%% file: main.tex
\documentclass[12pt,letterpaper]{article}
\usepackage[margin=1in]{geometry}
\usepackage{fontspec}
\usepackage{amsmath,amssymb,bm,mathtools}
\usepackage{graphicx,tikz}
\usetikzlibrary{arrows.meta,positioning}
\usepackage{setspace,booktabs,array,tabularx,longtable,multirow,makecell}
\usepackage{etoolbox}
\makeatletter
\patchcmd{\LT@output}{\vss}{\vskip 0pt plus 1fil minus 1pt}{}{}
\patchcmd{\LT@output}{\vss}{\vskip 0pt plus 1fil minus 1pt}{}{}
\makeatother
\usepackage{pdflscape,siunitx,enumitem,xcolor,url}
\usepackage[style=apa,backend=biber,natbib=true,sortcites=true]{biblatex}
\DeclareLanguageMapping{english}{english-apa}
\usepackage[hidelinks]{hyperref}
\usepackage{caption}
\newcolumntype{C}[1]{>{\centering\arraybackslash}p{#1}}
\newcolumntype{P}[1]{>{\raggedright\arraybackslash}p{#1}}
\newlength{\StudyTableWidth}
\usepackage{titlesec}
\titleformat{\section}{\centering\normalsize\bfseries}{}{0pt}{}
\titleformat{\subsection}{\normalsize\bfseries}{}{0pt}{}
\titleformat{\subsubsection}{\normalsize\bfseries\itshape}{}{0pt}{}
\titlespacing*{\section}{0pt}{14pt}{8pt}
\titlespacing*{\subsection}{0pt}{11pt}{5pt}
\hypersetup{pdftitle={Interpreting LLM Scores in Peer Review},pdfauthor={Wenqiang Qin; Yanjun Qian; Salma Srifi; Renlong Jie},bookmarksdepth=1}
\AtBeginBibliography{\setstretch{1.0}\setlength{\emergencystretch}{4em}\sloppy}
\makeatletter
\renewcommand{\@maketitle}{\begin{center}
{\normalsize\bfseries\@title\par}\vskip 1em
{\normalsize\lineskip .5em\begin{tabular}[t]{c}\@author\end{tabular}\par}
\end{center}\vskip .5em}
\makeatother
\begin{document}
\title{Interpreting LLM Scores in Peer Review: Alignment and Divergence Across Prompts, Domains, and Manuscript Features}
\author{Wenqiang Qin, Renlong Jie\thanks{Corresponding author: Renlong Jie. Email: \href{mailto:jierenlong@nwpu.edu.cn}{jierenlong@nwpu.edu.cn}.}, Yanjun Qian, Salma Srifi}
\date{}\maketitle
\begin{center}\small School of Management, Northwestern Polytechnical University, Xi'an, China\end{center}
\begin{abstract}
Large language model (LLM) scores are being studied as information for scholarly evaluation. Interpreting these scores requires evidence about agreement with human judgments and its variation across evaluation conditions. Guided by an argument-based approach to validation, we compare four LLMs on 7,083 public ICLR 2024 submissions under a standard review condition and examine ten prompt conditions using DeepSeek-V3.2. We distinguish average score divergence, paper-level absolute error, and rank correlation, then examine domain differences and adjusted manuscript-feature associations. For DeepSeek-V3.2, Area Chair guidance reduces mean rating divergence from 1.460 to 0.199 points and mean absolute error from 1.577 to 0.942, while rank correlation changes from 0.377 to 0.387. Historical references combined with downward scoring instructions yield mean divergence of 0.052 and rank correlation of 0.321. Domain differences remain after global calibration. Longer titles have positive adjusted associations with rating divergence, and formula density has negative associations across all four configurations. These findings connect score interpretation to the scoring procedure, evaluation context, and manuscript information represented in the comparison. They show why aggregate alignment, manuscript ordering, and correspondence within manuscript groups require distinct evidence when evaluating LLM-generated scholarly assessment information.
\end{abstract}

\section{Introduction}
Scores condense evaluations into information for comparing manuscripts. Large language models (LLMs) can produce these scores from manuscript text and a rubric. Interpreting their correspondence with human ratings requires examining score levels, differences between individual papers, and variation across research areas and manuscript characteristics. These questions concern how scholarly assessment information is represented and interpreted.

Recent research has examined how LLMs evaluate scholarly work. \citet{thelwall2025complex} discusses the evaluation of research quality as a complex text task. \citet{kousha2025societal} compare ChatGPT evaluations of societal impact cases with departmental expert averages and find variation across fields. \citet{wu2025novelty} combine human review information and LLM knowledge to predict methodological novelty. \citet{zhu2026language} identify model-dependent language bias in academic evaluation. Together, these studies show that model-generated assessment depends on the information supplied and the evaluation procedure.

The link between a numerical result and its intended interpretation provides our theoretical starting point. In an argument-based approach to validation, evidence is evaluated against the inferences required for a particular score interpretation or use \citep{kane2013validating}. Applied here, a close average score, a close score for each paper, and a similar ordering address different claims. Evidence about one property leaves the others to be examined. The research problem is therefore to establish how these properties change together across evaluation conditions.

Public OpenReview records link manuscripts to several human ratings \citep{tran2020open}. Meta-analytic evidence documents limited agreement among reviewers \citep{bornmann2010reliability}, and experimental evidence shows that author-identity information can affect recommendations \citep{tomkins2017reviewer}. These findings motivate examining model-human correspondence alongside the reliability and conditions of the human reference.

We study 7,083 public ICLR 2024 submissions with three to five human reviews per paper. We compare GPT-5 mini, Gemini 3 Flash Preview, Claude Sonnet 4.5, and DeepSeek-V3.2 under a standard review condition. We also evaluate ten prompt conditions using DeepSeek-V3.2. Mean divergence, mean absolute error (MAE), and Spearman correlation describe complementary agreement properties. Global calibration examines domain variation, and double/debiased machine learning (DML) estimates adjusted manuscript-feature associations.

The results reveal different responses across these properties. Area Chair guidance reduces mean divergence and MAE, with a small change in rank correlation. Historical references combined with downward scoring instructions give the closest mean and a lower rank correlation. Differences among domains persist after global calibration. Title length and formula density have recurring positive and negative adjusted associations with signed rating divergence, respectively.

The study connects scoring procedures, evaluation contexts, and manuscript information within one empirical design. It extends prompt comparisons in AI reviewing \citep{drori2024human} by examining score alignment, paper-level deviations, and ranking together. The domain and manuscript-feature analyses show where correspondence varies beneath aggregate agreement.

\section{Theoretical Background and Research Questions}
\subsection{Score Interpretation in Scholarly Evaluation}
Validity concerns whether evidence supports a proposed interpretation and the actions based on it. \citet{messick1995validity} links assessment evidence to score meaning and its consequences. \citet{kane2013validating} develops an argument-based approach that makes the required inferences and assumptions explicit. It also separates support for an interpretation from support for a consequential use. We apply this perspective to model-generated scholarly scores by specifying the claims that each empirical comparison addresses.

We use \textit{score correspondence} to describe the relationship between a model-generated rating and recorded human ratings of the same manuscript. \textit{Level agreement} concerns signed differences on a common scale. \textit{Paper-level agreement} concerns absolute distance from the human mean. \textit{Rank correspondence} concerns manuscript ordering. Each property supports a different interpretation of correspondence with recorded judgments.

Agreement and correlation address different properties of paired measurements \citep{bland1986agreement}. Positive and negative deviations can cancel in the mean, while similar rankings can coexist with systematic score offsets. A positive affine transformation changes location and spread while preserving ranks. We therefore assess mean divergence, MAE, and rank correlation together when comparing scoring instructions.

The reference also matters. \citet{kousha2025societal} uses departmental average impact scores. \citet{thelwall2025openalex} compare citation indicators against ChatGPT scores and departmental expert averages. \citet{zheng2026retraction} compare human and LLM classifications of critical scholarly tweets. Here, manuscript-level human means and pairwise human disagreement define the comparison.

\subsection{Scoring Procedures and Evaluation Contexts}
Prompt instructions are part of the scoring procedure. They specify the role, evaluation criteria, and rules for translating an assessment into a number. Instructions that request conservative ratings or provide historical score patterns can alter score levels. Changes in ordering depend on how the instructions affect distinctions among papers. This reasoning motivates examining several properties together when comparing prompts.

\citet{drori2024human} study cumulative GPT-4 reviewing instructions. \citet{thelwall2025complex} emphasizes systematic testing and repeated outputs. Work on model families and prompting further examines scoring procedures \citep{thelwall2026small}. Our paired design compares complete instruction bundles across ten DeepSeek conditions on the same submissions.

Evaluation contexts differ in manuscript evidence and review practices. Domains here are topic areas within one machine-learning conference. Overall calibration applies the same transformation to every paper, leaving domain-level correspondence to be examined. Human-review reliability also varies across studies and review conditions \citep{bornmann2010reliability}. Reporting within-domain human consistency helps interpret these comparisons.

\citet{zhu2026language} find opposite language preferences across two models evaluating humanities and social-science papers. Our domain comparison examines topic differences among real submissions after aggregate calibration, alongside human-reference consistency.

\subsection{Manuscript Information and Conditional Correspondence}
Document characteristics have an established place in research evaluation. \citet{kousha2024factors} review associations of titles, abstracts, document length, references, and author information with citations and expert quality judgments. Many relationships depend on the corpus and field. This makes manuscript features relevant to conditional evaluation, while keeping citation impact, expert judgment, and model-human divergence conceptually distinct.

\citet{wu2025novelty} combine human review statements and model-derived method summaries for novelty prediction. Research on long contexts and controlled presentation changes examines model responses to supplied text \citep{liu-etal-2024-lost,du2025titletrap,yang2026presentation}. We examine which manuscript characteristics are associated with differences between model and human ratings after adjustment. This connects feature-based research assessment to correspondence between evaluation procedures.

Titles and formulas carry substantive content and presentation choices. We compare observed manuscript groups after adjustment for covariates and a full-text predictor of human ratings. Overlap weighting emphasizes manuscripts with comparable adjustment variables. The contrast describes signed divergence conditional on this information. A negative contrast indicates lower divergence in the compared group. Its implication for absolute agreement depends on the score offset and individual errors. Recurring signs across models identify groups for further validation.

\subsection{Research Questions and Evidence Map}
We examine three questions derived from these interpretive requirements:
\begin{enumerate}[label=RQ\arabic*:,leftmargin=3em]
\item How do score levels, paper-level agreement, and rank correspondence vary across prompt conditions, relative to disagreement among human reviewers?
\item How does model-human correspondence vary across domains after aggregate scale alignment, and across human-rating groups?
\item Which adjusted associations between manuscript characteristics and signed rating divergence recur across model configurations?
\end{enumerate}
Table~\ref{tab:validation_map} maps the questions to their evidence. Table~\ref{tab:related_evidence} positions the study relative to the closest research. The framework guides interpretation of the analyses.

\begin{table}[!ht]
\centering\small
\caption{Score Interpretations and Corresponding Evidence}\label{tab:validation_map}
\begin{tabularx}{\linewidth}{@{}P{0.23\linewidth}XX@{}}
\toprule
Interpretive question & Evidence in this study & What the evidence addresses \\
\midrule
How closely does a procedure reproduce recorded assessments? & Mean divergence, MAE, rank correlation, and reviewer disagreement across prompts & Average levels, individual-paper distances, and ordering are evaluated separately \\
Does overall correspondence extend across contexts? & Domain means and ranks after global calibration, with human ICC & Remaining contextual variation after aggregate scale alignment \\
Does correspondence vary with manuscript information? & Adjusted feature contrasts and sensitivity specifications across models & Conditional signed differences between observed manuscript groups \\
\bottomrule
\end{tabularx}
\end{table}

\begin{table}[!ht]
\centering\small
\caption{Related Evaluation Research and the Present Study}\label{tab:related_evidence}
\begin{tabularx}{\linewidth}{@{}P{0.24\linewidth}XX@{}}
\toprule
Research & Main focus & Relation to the present study \\
\midrule
\citet{drori2024human} & Cumulative GPT-4 reviewing instructions & Paired comparison of instruction bundles on levels, absolute deviations, and ranks \\
\citet{kousha2024factors} & Document features associated with citations or research quality & Feature associations with signed differences between model and human evaluations \\
\citet{kousha2025societal} & ChatGPT impact-case ratings and departmental expert averages & Manuscript-level human references and joint agreement measures \\
\citet{wu2025novelty} & Novelty prediction using human and LLM information & Interpretation of scores across procedures and manuscript groups \\
\citet{zhu2026language} & Model-dependent language bias in academic evaluation & Prompt, topic-area, and manuscript-feature variation in score correspondence \\
\citet{xie2026confidence} & Reviewer confidence and review breadth/depth & Human-reference variation considered alongside model-human agreement \\
\bottomrule
\end{tabularx}
\end{table}

\section{Research Design}
\label{sec:methods}
\begin{figure}[!ht]
\centering
\includegraphics[width=\linewidth]{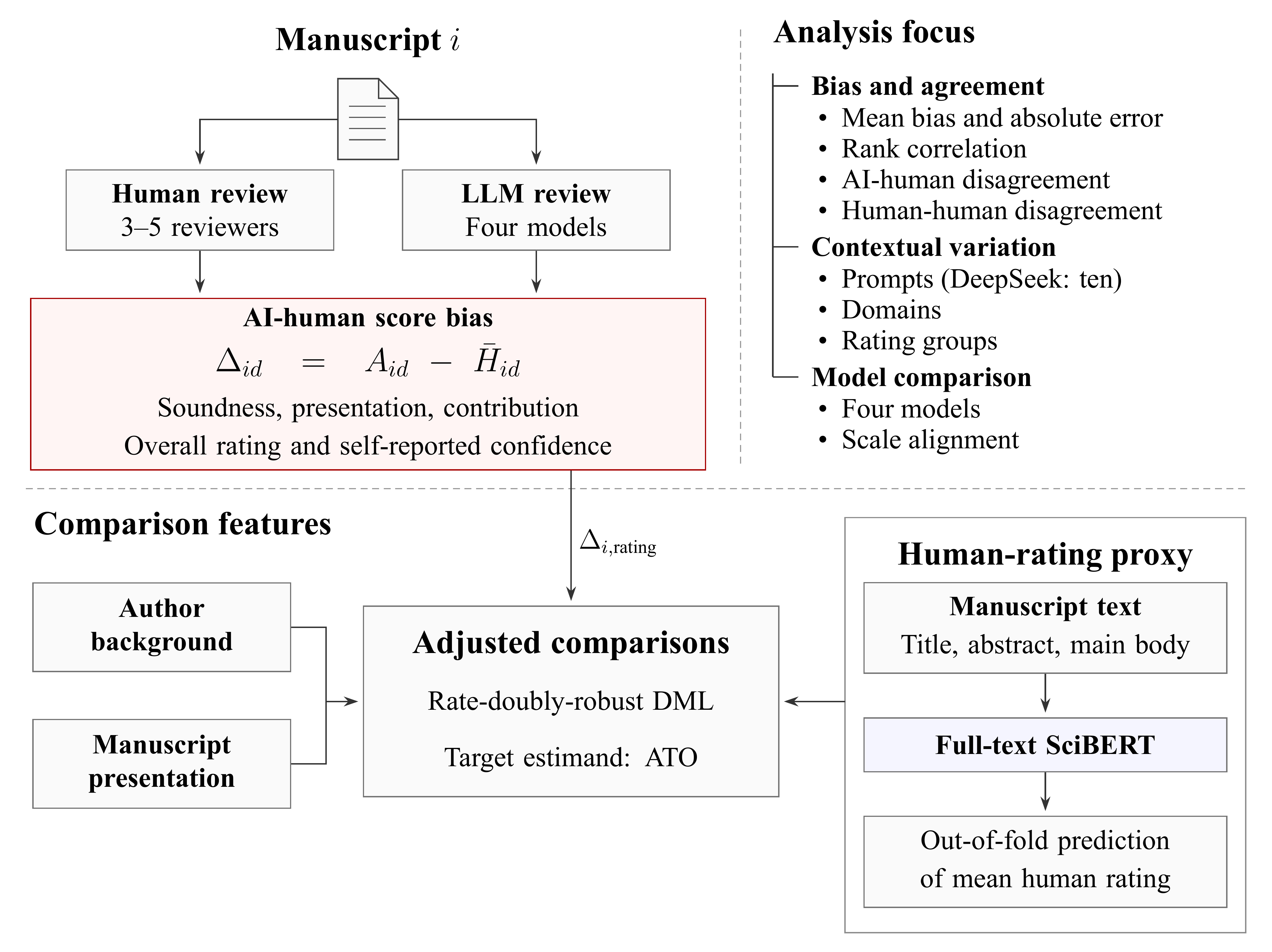}
\caption{Research Framework. Human scores are averaged across three to five reviewers and compared with each of four LLMs. The analysis covers score bias, absolute error, rank correlation, reviewer disagreement, and variation across ten DeepSeek prompt conditions, domains, and human-rating groups. Confidence records self-reported certainty. The lower panels use rate-doubly-robust DML and an out-of-fold SciBERT predictor of mean human ratings to estimate adjusted group differences using the ATO estimand. The arrows show the analysis workflow. The predictor summarizes manuscript information associated with human ratings.}
\label{fig:flowchart}
\end{figure}

The design connects paired manuscript scores to the three research questions. Prompt comparisons examine the scoring procedure, domain comparisons examine contextual correspondence, and adjusted feature comparisons examine differences among observed manuscript groups. Figure~\ref{fig:flowchart} summarizes this workflow.

\subsection{Data, Human Reference, and Model Configurations}
The study uses retrospective public records as a setting for evaluating model-generated assessment information. We obtained approximately 7,400 public ICLR 2024 submission PDFs and their review records from OpenReview. We retained papers with three to five valid human reviews, yielding 7,083 papers. GROBID parsed the PDFs into structured text. We used this output to extract titles, abstracts, manuscript bodies, formulas, figures, and tables \citep{lopez2009grobid,GROBID}. Human reviews included soundness, presentation, contribution, overall rating, and reviewer confidence. The first four concern manuscript evaluation. Confidence records the reviewer's self-assessment.

We compared GPT-5 mini, Gemini 3 Flash Preview, Claude Sonnet 4.5, and DeepSeek-V3.2 under the standard review condition. The protocol specified requested temperature $0.1$ and top-$p$ $0.95$. Appendices~\ref{app:prompt_design}-\ref{app:prompt_exact} describe the prompt design, extended results, diagnostics, and implementation.

We used each paper's mean human score as the reference for each dimension. Throughout the study, \textit{bias} denotes directional AI-human divergence. We linked author characteristics to manuscript records after data collection. Author variables enter secondary comparisons and adjustment. The country-based variable labeled Native English is a proxy based on the listed country group, with the coding shown in Table~\ref{tab:variable_definitions666}.

\subsection{Measurement of Evaluation Properties}
Let $A_{id}$ be the AI score for paper $i$ and dimension $d$, $H_{idr}$ the score of human reviewer $r$, $\bar H_{id}$ its human mean, and $m_{id}$ the number of valid reviews. Directional divergence and absolute error are
\[
\Delta_{id}=A_{id}-\bar H_{id},\qquad
\operatorname{MAE}_d=N_d^{-1}\sum_i|\Delta_{id}|.
\]
We also measure disagreement on an individual-reviewer basis:
\[
D^{AH}_d=\frac{1}{N_d}\sum_i\frac{1}{m_{id}}\sum_r|A_{id}-H_{idr}|,
\]
\[
D^{HH}_d=\frac{1}{N_d}\sum_i\frac{2}{m_{id}(m_{id}-1)}
\sum_{r<r'}|H_{idr}-H_{idr'}|.
\]
Here, $N_d$ is the number of papers with valid scores on dimension $d$. $D^{AH}_d$ averages disagreement between the AI score and each human score. $D^{HH}_d$ averages disagreement among all pairs of human reviewers. Comparing them places AI-human disagreement on the same individual-reviewer basis as human-human disagreement. We measured relative discrimination using Spearman correlation between $A_{id}$ and $\bar H_{id}$. We applied Benjamini-Hochberg adjustment within the reported test families \citep{benjamini1995fdr}.

\subsection{Prompt Comparisons and Contextual Heterogeneity}
We tested ten DeepSeek conditions: five cumulative structured prompts and five tone/depth variants. V1 supplied the paper and review form. V2 added reviewer guidelines, and V3 added conduct and ethics guidance. V4 added Area Chair selectivity, including choosing the lower rating when undecided. V5 added historical score patterns and a scoring procedure: form a preliminary rating, then usually lower it by about one point, or up to two for noticeable weaknesses. The tone variants used the same numeric scales with a different section order, including Questions and Ethics flag and details. V4 and V5 each changed several instructions. Appendices~\ref{app:prompt_design} and \ref{app:prompt_exact} document these instruction bundles and their implementation.

We evaluated the same papers under each prompt. We used Friedman tests for bias and absolute error, Cochran's $Q$ for directional rates, and permutation tests for rank-correlation differences. False discovery rate (FDR) adjustment was applied within each dimension. The standard condition had the highest observed DeepSeek rank correlation and was selected for the cross-model analyses. Selection and evaluation used the same papers, so the reported comparison describes performance within this sample.

We assessed the human intraclass correlation coefficient (ICC(1,$k$)), mean divergence, MAE, and rank correlation by domain. Human consistency was examined in the 12 primary areas with at least 200 papers (Figure~\ref{fig:area_human_icc}). The seven-area model comparison covered a subset of 3,684 papers from the 7,083 submissions. We also calibrated raw model scores to separate differences in score location and scale from remaining domain variation:
\[
A'=aA+b,\qquad a=\frac{\sigma(H)}{\sigma(A)},\quad b=\mu(H)-a\mu(A).
\]
For nondegenerate scores, $a>0$, so this transformation preserves ranks. We estimated the parameters from the analyzed sample. The overall mean and standard deviation are aligned by construction. The transformation is a within-sample diagnostic for remaining domain differences. Area-model cell analyses used sample-size weights, nested model comparisons, a one-degree-of-freedom non-additivity test, and wild-bootstrap inference. This analysis decomposes score-scale and domain differences across 28 area-model cells. The four configurations evaluate overlapping sets of papers within each area. We treat the cell-level inferential decomposition as exploratory and emphasize the reported area-specific means and rank correlations.

We rounded each paper's mean human rating to the nearest integer and limited it to the 1-10 scale. We then defined low (1-4), middle (5-6), and high (7-10) groups. Figure~\ref{fig:rating_bins_bias_corr_bars} uses the labels Low, Middle, and High for these groups. Their membership was determined by ratings. For the human mean $H_i=\bar H_{i,\mathrm{rating}}$, the identity $\mathbb E[\Delta_i\mid H_i=h]=\mathbb E[A_i\mid H_i=h]-h$ shows that conditional divergence depends on the reference score itself. We use these strata to describe the sample, with mathematical coupling and within-stratum range restriction considered in interpretation. We tested bias differences using analysis of variance (ANOVA) and Kruskal-Wallis tests. Correlation heterogeneity was assessed using Fisher-$z$-based Cochran tests and $I^2$.

\subsection{Adjusted Manuscript-Feature Comparisons}
We first used ordinary least squares (OLS) to relate overall rating divergence $\Delta_i$ to author and manuscript characteristics (Table~\ref{tab:variable_definitions666}). We calculated heteroskedasticity-consistent standard errors using the HC3 estimator. The ex-ante specification included author and manuscript variables. The ex-post specification added human-rating dispersion and reviewer count. Missing continuous values were imputed as zero in the OLS analysis.

\par
\begingroup
\small\setstretch{1.05}
\setlength{\tabcolsep}{3pt}
\renewcommand{\arraystretch}{1}
\setlength{\StudyTableWidth}{\dimexpr\linewidth-4\tabcolsep-3\arrayrulewidth\relax}
\begin{longtable}{P{0.28\StudyTableWidth}P{0.72\StudyTableWidth}}
\caption{Variable Definitions}\label{tab:variable_definitions666}\\
\toprule
\textbf{Variable} & \textbf{Description} \\
\midrule
\endfirsthead
\multicolumn{2}{P{\dimexpr\linewidth-2\tabcolsep-2\arrayrulewidth\relax}}{\textit{Table~\thetable\ (continued)}} \\
\textbf{Variable} & \textbf{Description} \\
\midrule
\endhead
\endfoot
\endlastfoot
\multicolumn{2}{P{\dimexpr\linewidth-2\tabcolsep-2\arrayrulewidth\relax}}{\textit{Author characteristics}} \\
Industry involvement & Share of authors affiliated with industry or non-academic institutions \\
Top institution & Share of authors affiliated with top AI institutions \\
Female authors & Share of female authors \\
First author student & Indicator for whether the first author is identified as a student \\
Native English & Share of authors from English-speaking countries \\
N authors & Number of authors listed for the paper \\
\multicolumn{2}{P{\dimexpr\linewidth-2\tabcolsep-2\arrayrulewidth\relax}}{\textit{Paper characteristics}} \\
Formula density & Number of formulas per 1,000 main-text characters \\
Fig.\ \& Tab.\ density & Number of figures and tables per 1,000 main-text characters \\
Table info density & Number of extracted table elements per table character length \\
Citation density & Number of references per 1,000 main-text characters \\
Appendix ratio & Appendix character length divided by total paper character length \\
Log body length & Log-transformed main-text character length \\
Title length & Character length of the title \\
Avg formula length & Average character length of formulas \\
Avg sentence length & Average character length of main-text sentences \\
Abstract length & Character length of the abstract \\
\multicolumn{2}{P{\dimexpr\linewidth-2\tabcolsep-2\arrayrulewidth\relax}}{\textit{Notes:} Top AI institutions comprise the top 50 academic institutions
listed by AI Rankings and the top 50 AI companies by market
capitalization. Institutional classifications use AI Rankings and CompaniesMarketCap. English-speaking countries include US, GB,
AU, NZ, and IE.} \\
\bottomrule
\end{longtable}
\endgroup

DML added adjustment for a hierarchical full-text SciBERT predictor of mean human ratings \citep{beltagy2019scibert,chernozhukov2018double}. This predictor summarizes textual information associated with human evaluation. Its target is the mean human rating. We obtained $\hat H_i$ through five-fold out-of-fold prediction at the paper level, using the title, abstract, and main body. We excluded appendices and references. Appendix~\ref{app:implementation} describes the architecture and estimation score.

We compared each characteristic in turn. Binary variables retained their coding. For other variables, we compared the upper and lower terciles and excluded the middle tercile. Author-characteristic comparisons adjusted for the proxy, manuscript variables, review-condition controls, and primary-area indicators. Manuscript-characteristic comparisons adjusted for the proxy, author variables, review-condition controls, and area indicators. Thus, each comparison adjusted for characteristics in the other block.

Let $T$ identify the high group for the characteristic being compared, and let $X$ denote its adjustment variables. The propensity $e(X)=\Pr(T=1\mid X)$ is the probability of belonging to the high group given $X$. The outcome function $m_t(X)=\mathbb E[\Delta\mid T=t,X]$ is expected rating divergence for group $t$. The average treatment effect for the overlap population (ATO) uses overlap weights \citep{li2018balancing}:
\[
\theta_{\mathrm{ATO}}=
\frac{\mathbb E[e(X)\{1-e(X)\}\{m_1(X)-m_0(X)\}]}
{\mathbb E[e(X)\{1-e(X)\}]}.
\]
The propensity multiplied by its complement gives greater emphasis to papers whose observed characteristics are compatible with both groups. Each manuscript characteristic defines its own overlap population, so comparisons across rows concern different populations. We used the ATO-specific rate-doubly-robust estimator to account for the dependence of these weights on the propensity \citep{wang2025rate}. Ensemble nuisance models used five-fold paper-level cross-fitting, repeated across ten partitions. We reported median estimates with paper-clustered influence-function inference and adjustment for variation across partitions. FDR adjustment was applied separately within each model and author/manuscript family. Appendix~\ref{app:implementation} gives the estimating score, training settings, and clustered inference.

We interpret $\theta_{\mathrm{ATO}}$ as an adjusted difference in rating divergence between naturally occurring high and low groups. The contrast is expressed in rating points on the 1-10 scale. Adjustment conditions on the observed covariates and human-rating predictor used in each specification. Titles and formulas carry both substantive information and presentation choices, which jointly enter these group comparisons.

\section{Results}
\label{sec:results}
\subsection{Score Divergence Relative to Human Disagreement}
Table~\ref{tab:core_bias} reports the V1 DeepSeek baseline. AI-human disagreement exceeds human-human disagreement for soundness, contribution, and rating, with ratios of 1.11, 1.31, and 1.31, respectively ($q<0.001$). Presentation disagreement is comparable to human disagreement (ratio 1.00, $q=0.470$). Mean divergence is positive on every dimension. Overall rating divergence is 1.460 points, with a standardized effect of 1.26. Confidence scores show less AI-human disagreement than human-human disagreement, describing agreement in reported certainty.
\par
\begingroup
\small\setstretch{1.05}
\setlength{\tabcolsep}{3pt}
\renewcommand{\arraystretch}{1}
\setlength{\StudyTableWidth}{\dimexpr\linewidth-14\tabcolsep-8\arrayrulewidth\relax}
\begin{longtable}{P{0.18\StudyTableWidth}C{0.11\StudyTableWidth}C{0.11\StudyTableWidth}C{0.1\StudyTableWidth}C{0.13\StudyTableWidth}C{0.13\StudyTableWidth}C{0.24\StudyTableWidth}}
\caption{DeepSeek V1 Baseline: AI-Human Divergence and Human Disagreement}\label{tab:core_bias}\\
\toprule
\textbf{Dimension} & \textbf{$D^{AH}_{d}$} & \textbf{$D^{HH}_{d}$} & \textbf{Ratio} & \textbf{$q$ (FDR)} & \textbf{Mean bias} & \textbf{Bias Cohen's $d$} \\
\midrule
\endfirsthead
\multicolumn{7}{P{\dimexpr\linewidth-2\tabcolsep-2\arrayrulewidth\relax}}{\textit{Table~\thetable\ (continued)}} \\
\textbf{Dimension} & \textbf{$D^{AH}_{d}$} & \textbf{$D^{HH}_{d}$} & \textbf{Ratio} & \textbf{$q$ (FDR)} & \textbf{Mean bias} & \textbf{Bias Cohen's $d$} \\
\midrule
\endhead
\endfoot
\endlastfoot
Soundness & 0.705 & 0.635 & 1.11 & $<0.001$ & 0.559$^{***}$ & 1.33 \\
Presentation & 0.685 & 0.684 & 1.00 & 0.470 & 0.422$^{***}$ & 0.89 \\
Contribution & 0.830 & 0.633 & 1.31 & $<0.001$ & 0.691$^{***}$ & 1.45 \\
Rating & 1.831 & 1.397 & 1.31 & $<0.001$ & 1.460$^{***}$ & 1.26 \\
Confidence & 0.614 & 0.783 & 0.78 & 1.000 & 0.358$^{***}$ & 0.73 \\
\multicolumn{7}{P{\dimexpr\linewidth-2\tabcolsep-2\arrayrulewidth\relax}}{\textit{Note:} This table reports DeepSeek-V3.2 under V1. Mean bias is $AI-\bar H$.  \(q\) is the BH-adjusted one-sided \(p\)-value for
\(D^{AH}_{d}>D^{HH}_{d}\). Stars report BH-adjusted two-sided tests of Mean
Bias against zero. \(^{***}p_{\mathrm{FDR}}<0.001\).} \\
\bottomrule
\end{longtable}
\endgroup

\subsection{Prompt Conditions and Distinct Agreement Properties}
Table~\ref{tab:rating_integrated_sd} compares the ten DeepSeek conditions on the same 7,083 papers. V1-V3 show large positive mean divergence and upward-scoring rates above 85\%. V4 has the lowest MAE, 0.942, and mean divergence of 0.199, compared with 1.577 and 1.460 for V1. Its rank correlation is 0.387, close to V1's 0.377. The Area Chair instructions substantially improve score alignment, alongside a small change in rank agreement.

V5 has the smallest mean divergence, 0.052, and the lowest rank correlation, 0.321. Its MAE is 0.961. These results reflect the combined historical reference values and downward scoring instructions in V5. The standard condition has the highest rank correlation, 0.423, and a larger MAE of 1.464. The critical condition lowers MAE to 1.115 with a correlation of 0.403. Domain-expert and quick instructions yield correlations of 0.415 and 0.407, with MAE near 1.49.
\par
\begingroup
\small\setstretch{1.05}
\setlength{\tabcolsep}{3pt}
\renewcommand{\arraystretch}{1}
\setlength{\StudyTableWidth}{\dimexpr\linewidth-14\tabcolsep-8\arrayrulewidth\relax}
\begin{longtable}{P{0.17\StudyTableWidth}C{0.18\StudyTableWidth}C{0.18\StudyTableWidth}C{0.1\StudyTableWidth}C{0.1\StudyTableWidth}C{0.1\StudyTableWidth}C{0.17\StudyTableWidth}}
\caption{DeepSeek Prompt Comparison on Overall Rating ($N=7{,}083$)}\label{tab:rating_integrated_sd}\\
\toprule
\textbf{Prompt} & \textbf{Mean bias (SD)} & \textbf{MAE (SD)} & \textbf{Over (\%)} & \textbf{Under (\%)} & \textbf{Exact (\%)} & \textbf{Spearman $\rho$} \\
\midrule
\endfirsthead
\multicolumn{7}{P{\dimexpr\linewidth-2\tabcolsep-2\arrayrulewidth\relax}}{\textit{Table~\thetable\ (continued)}} \\
\textbf{Prompt} & \textbf{Mean bias (SD)} & \textbf{MAE (SD)} & \textbf{Over (\%)} & \textbf{Under (\%)} & \textbf{Exact (\%)} & \textbf{Spearman $\rho$} \\
\midrule
\endhead
\endfoot
\endlastfoot
V1 & 1.460\newline (1.157) & 1.577\newline (0.992) & 87.05 & 8.50 & 4.45 & 0.377 \\
V2 & 1.397\newline (1.160) & 1.526\newline (0.985) & 85.92 & 9.33 & 4.74 & 0.392 \\
V3 & 1.399\newline (1.148) & 1.528\newline (0.970) & 86.04 & 9.33 & 4.63 & 0.399 \\
V4 & 0.199\newline (1.153) & 0.942\newline (0.693) & 52.82 & 38.50 & 8.68 & 0.387 \\
V5 & 0.052\newline (1.183) & 0.961\newline (0.693) & 48.20 & 43.80 & 8.01 & 0.321 \\
constructive & 1.295\newline (1.148) & 1.449\newline (0.945) & 83.74 & 10.94 & 5.32 & 0.389 \\
domain expert & 1.362\newline (1.134) & 1.495\newline (0.951) & 85.35 & 9.70 & 4.96 & 0.415 \\
critical & 0.748\newline (1.146) & 1.115\newline (0.793) & 70.01 & 22.31 & 7.68 & 0.403 \\
quick & 1.359\newline (1.136) & 1.490\newline (0.956) & 85.61 & 9.64 & 4.74 & 0.407 \\
standard & 1.330\newline (1.127) & 1.464\newline (0.946) & 84.94 & 9.83 & 5.24 & 0.423 \\
\multicolumn{7}{P{\dimexpr\linewidth-2\tabcolsep-2\arrayrulewidth\relax}}{\textit{Note:} All conditions use DeepSeek-V3.2. Bias is the model score minus the mean human score. Exact denotes identical numeric scores. Its frequency depends on comparing integer model outputs with averaged human scores. Table~\ref{tab:omnibus_prompt_differences} reports omnibus tests.} \\
\bottomrule
\end{longtable}
\endgroup

Table~\ref{tab:prompt_sensitivity_compact} summarizes variation across dimensions, with omnibus tests in Table~\ref{tab:omnibus_prompt_differences}. Prompt variation is greater for manuscript evaluation than for self-reported confidence.
\par
\begingroup
\small\setstretch{1.05}
\setlength{\tabcolsep}{3pt}
\renewcommand{\arraystretch}{1}
\setlength{\StudyTableWidth}{\dimexpr\linewidth-8\tabcolsep-5\arrayrulewidth\relax}
\begin{longtable}{P{0.28\StudyTableWidth}C{0.22\StudyTableWidth}C{0.22\StudyTableWidth}C{0.28\StudyTableWidth}}
\caption{Prompt-Dependent Ranges Across Evaluation Dimensions ($N=7{,}083$)}\label{tab:prompt_sensitivity_compact}\\
\toprule
\textbf{Dimension} & \textbf{Bias range} & \textbf{MAE range} & \textbf{Spearman $\rho$ range} \\
\midrule
\endfirsthead
\multicolumn{4}{P{\dimexpr\linewidth-2\tabcolsep-2\arrayrulewidth\relax}}{\textit{Table~\thetable\ (continued)}} \\
\textbf{Dimension} & \textbf{Bias range} & \textbf{MAE range} & \textbf{Spearman $\rho$ range} \\
\midrule
\endhead
\endfoot
\endlastfoot
Soundness & 0.635 & 0.244 & 0.119 \\
Presentation & 0.334 & 0.122 & 0.088 \\
Contribution & 0.522 & 0.342 & 0.091 \\
Rating & 1.408 & 0.635 & 0.102 \\
Confidence & 0.027 & 0.004 & 0.053 \\
\multicolumn{4}{P{\dimexpr\linewidth-2\tabcolsep-2\arrayrulewidth\relax}}{Note: Entries are maxima minus minima across ten DeepSeek conditions. Soundness, Presentation, and Contribution use 1-4 scales, Rating 1-10, and Confidence 1-5. The ranges describe variation on each dimension's original scale. Table~\ref{tab:omnibus_prompt_differences} reports omnibus tests.} \\*
\bottomrule
\end{longtable}
\endgroup

The omnibus tests identify differences in bias, MAE, and correlation across prompt conditions. For overall rating, the ranges are 1.408 for bias, 0.635 for MAE, and 0.102 for correlation. Evaluation properties thus respond differently to the instruction bundles.

\subsection{Correspondence Across Domains and Rating Groups}
Human rating consistency varies across the 12 domains with at least 200 papers (Figure~\ref{fig:area_human_icc}). Paper-level absolute human disagreement also varies across all five dimensions (Table~\ref{tab:kruskal_wallis_all_dims}). ICC is 0.725 for Generative Models and 0.720 for Reinforcement Learning, compared with 0.601 for Societal AI Ethics and 0.614 for Datasets and Benchmarks. Domain comparisons therefore need to consider variation in the human reference alongside model performance.

\begin{figure}[htbp]
\centering
\includegraphics[width=0.9\linewidth]{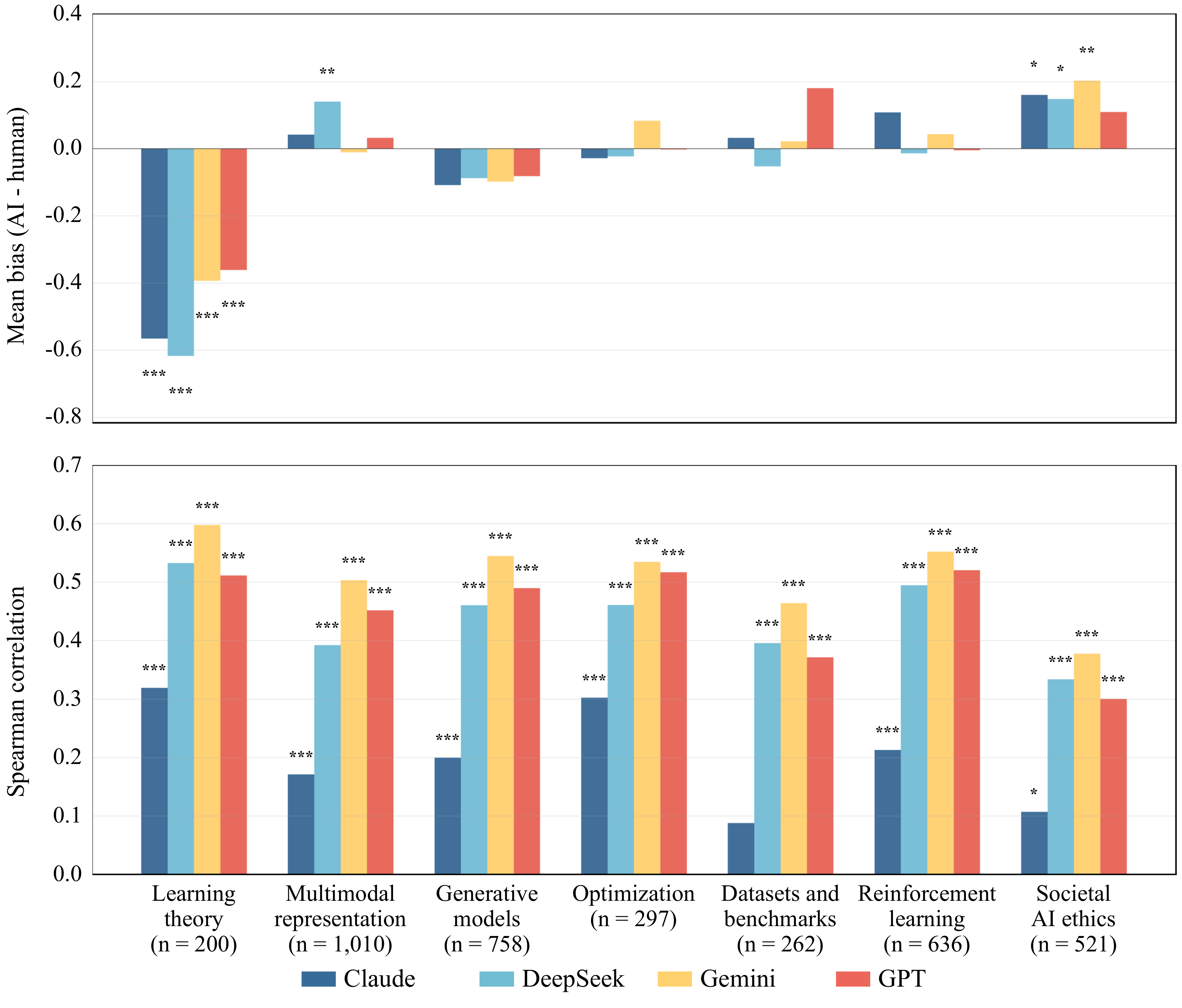}
\caption{Mean Bias and Rank Consistency Across Domains. The top panel reports mean bias after global affine calibration. The bottom reports Spearman rank consistency for four large language models.}
\label{fig:model_domain_4models}
\end{figure}

Figure~\ref{fig:model_domain_4models} shows model results after global affine calibration. Generative Models and Reinforcement Learning have small absolute mean divergence, at most 0.11. All four models show negative divergence in Learning Theory, ranging from $-0.36$ to $-0.62$, alongside relatively strong rank agreement. Societal AI Ethics shows positive divergence and weaker rank agreement. These domain patterns persist after correcting global score location and scale.

Table~\ref{tab:anova_rating_full} reports the full area-model cell analysis. The cell-level decomposition summarizes area and model components. Its inferential results are exploratory because the model cells share manuscripts within areas. Rank correlations are unchanged by the positive affine transformation. Global score calibration therefore leaves domain differences in divergence. Table~\ref{tab:area_struct_stats_compact_sd} describes the fields' manuscript characteristics, including formula and appendix use.

\begin{figure}[htbp]
\centering
\includegraphics[width=0.9\linewidth]{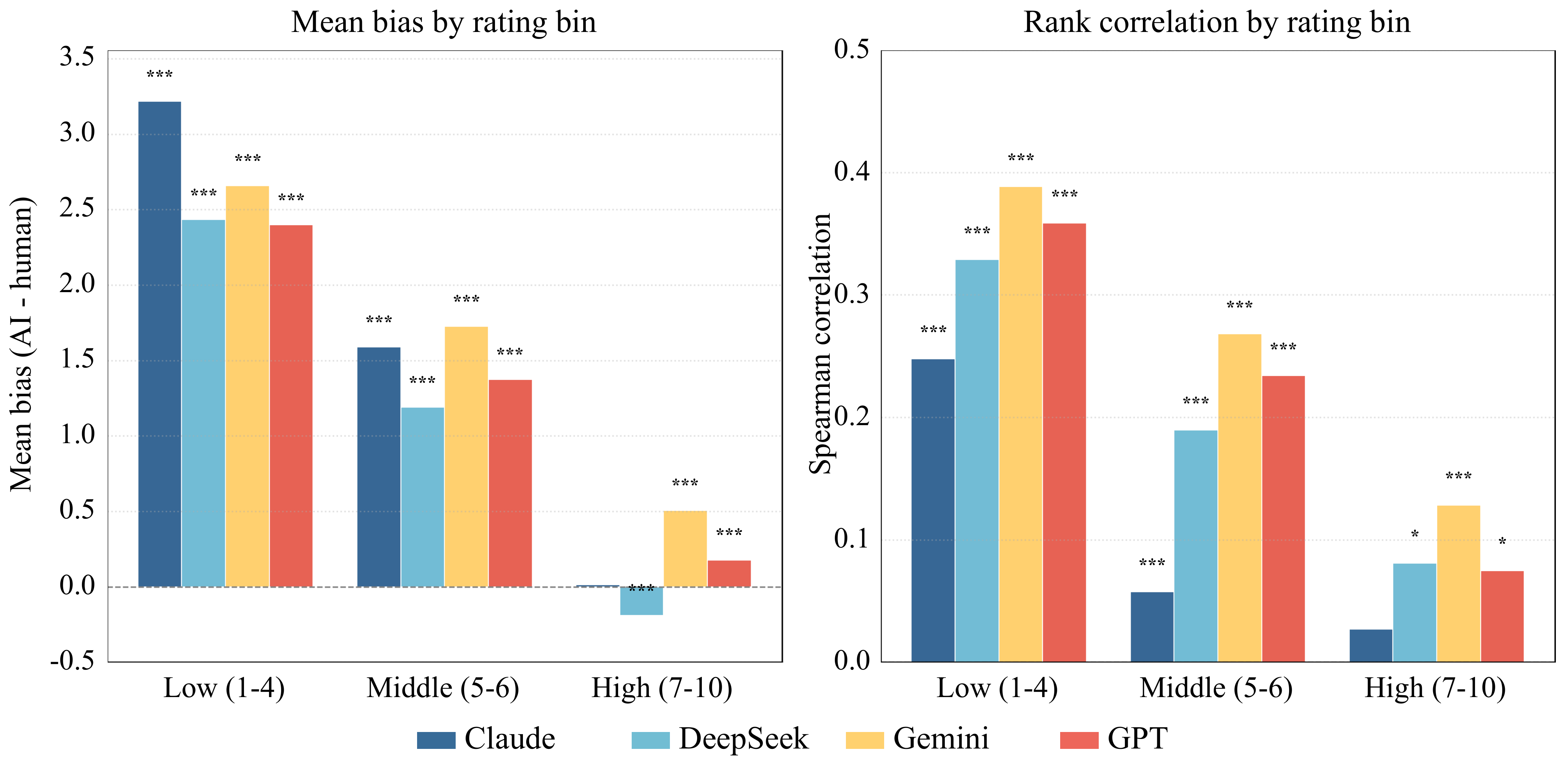}
\caption{Mean Bias and Spearman Correlation Across Score Bins. Bins are defined by rounded human ratings. The left panel shows mean bias (AI $-$ Human), and the right panel shows within-bin Spearman correlation. Asterisks indicate that the corresponding statistical test remains significant after FDR correction.}
\label{fig:rating_bins_bias_corr_bars}
\end{figure}

Figure~\ref{fig:rating_bins_bias_corr_bars} shows results by human-rating group. All four configurations have positive divergence of approximately 2.40-3.22 points in the low group and 1.19-1.73 in the middle group. Model-specific directions differ in the high group. Within-group rank correlations decline toward the high group. Table~\ref{tab:rating_bin_full} reports heterogeneity tests. Human-score standard deviations also decline, from 0.603 in the low group to 0.555 in the middle group and 0.469 in the high group (Table~\ref{tab:rating_bin_variance_summary}). Group membership uses the human score that also enters rating divergence. The group patterns therefore reflect conditional score comparisons, mathematical coupling, and range restriction alongside model performance.

\subsection{Manuscript-Feature Associations Across Models}
The OLS models explain different shares of rating divergence. $R^2$ is 0.339 for Claude, 0.209 for DeepSeek, 0.092 for Gemini, and 0.106 for GPT. Title length is positively associated with divergence in all four configurations. Formula density, appendix ratio, and abstract length are negatively associated. Adding human-rating dispersion and reviewer count increases $R^2$ by at most 0.002. Tables~\ref{tab:global_delta_ols_coef} and \ref{tab:global_delta_ols_filled_coef_B} report the full specifications.

\begin{figure}[htbp]
\centering
\includegraphics[width=\linewidth]
{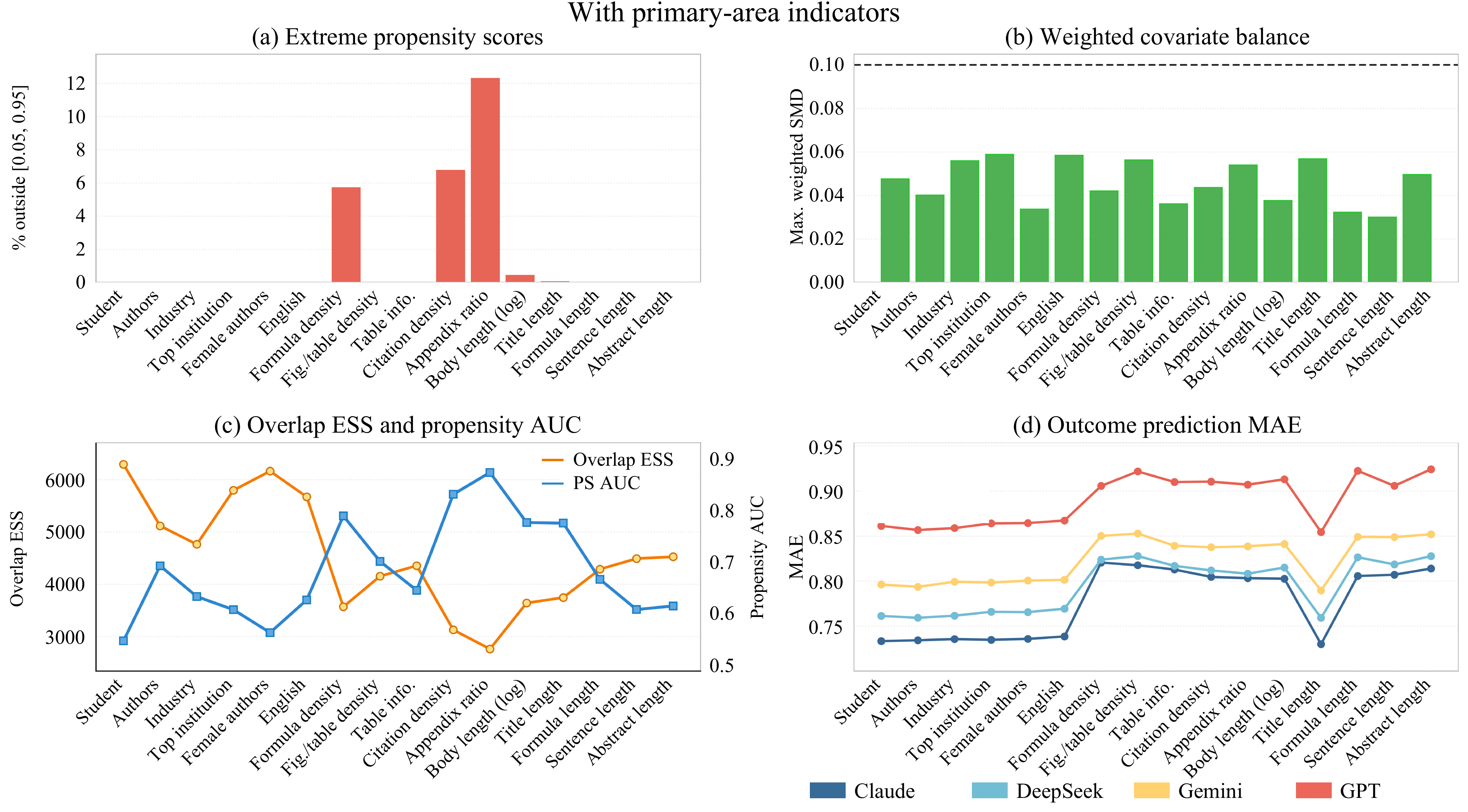}
\caption{Diagnostics for the Primary Overlap-Contrast Specification. Panel (a) reports propensity estimates outside $[0.05,0.95]$. Panel (b) reports the maximum absolute weighted standardized mean difference (SMD). Panel (c) reports effective sample size (ESS) and the area under the receiver operating characteristic curve (AUC) for the propensity model. Panel (d) reports cross-fitted outcome MAE. The panels describe observed overlap, covariate balance, and predictive accuracy.}
\label{fig:rdr_dml_diagnostics_with_area}
\end{figure}

The full-text SciBERT proxy predicts mean human rating with out-of-fold MAE 0.790, $R^2=0.369$, Pearson $r=0.614$, and Spearman $\rho=0.607$. After overlap weighting, maximum absolute standardized mean differences are below 0.06 for every compared characteristic. Figure~\ref{fig:rdr_dml_diagnostics_with_area} shows these diagnostics by characteristic. Appendix ratio, citation density, and formula density show greater separation in propensity scores, yet their weighted covariate differences remain below 0.06. Effective sample sizes also vary because each comparison concerns a different overlap population.
\par

Table~\ref{tab:rdr_dml_ato_main} reports the adjusted ATO contrasts. Title-length estimates are positive: 0.913, 0.828, 0.742, and 0.755 for Claude, DeepSeek, Gemini, and GPT. Formula-density estimates are negative: $-0.097$, $-0.169$, $-0.139$, and $-0.291$. All eight contrasts remain significant after FDR adjustment. For title length, these values compare the upper and lower terciles of title character counts. Formula-density values compare the corresponding terciles of formulas per 1,000 main-text characters. The differences are conditional associations in each characteristic-specific overlap population.
\par
\begingroup
\small\setstretch{1.05}
\setlength{\tabcolsep}{3pt}
\renewcommand{\arraystretch}{1}
\setlength{\StudyTableWidth}{\dimexpr\linewidth-10\tabcolsep-6\arrayrulewidth\relax}
\begin{longtable}{P{0.26\StudyTableWidth}C{0.185\StudyTableWidth}C{0.185\StudyTableWidth}C{0.185\StudyTableWidth}C{0.185\StudyTableWidth}}
\caption{Adjusted Overlap Contrasts in AI-Human Rating Divergence}\label{tab:rdr_dml_ato_main}\\
\toprule
\textbf{Variable} & \textbf{Claude\newline Contrast (SE)} & \textbf{DeepSeek\newline Contrast (SE)} & \textbf{Gemini\newline Contrast (SE)} & \textbf{GPT\newline Contrast (SE)} \\
\midrule
\endfirsthead
\multicolumn{5}{P{\dimexpr\linewidth-2\tabcolsep-2\arrayrulewidth\relax}}{\textit{Table~\thetable\ (continued)}} \\
\textbf{Variable} & \textbf{Claude\newline Contrast (SE)} & \textbf{DeepSeek\newline Contrast (SE)} & \textbf{Gemini\newline Contrast (SE)} & \textbf{GPT\newline Contrast (SE)} \\
\midrule
\endhead
\endfoot
\endlastfoot
\textit{Author characteristics} &  &  &  \\
First author student & 0.040\newline (0.024) & 0.013\newline (0.024) & -0.012\newline (0.026) & -0.029\newline (0.028) \\
N authors & -0.079**\newline (0.027) & -0.057\newline (0.027) & -0.046\newline (0.029) & -0.091**\newline (0.031) \\
Industry involvement & -0.077**\newline (0.027) & 0.004\newline (0.028) & 0.045\newline (0.029) & 0.104**\newline (0.031) \\
Top institution & -0.089**\newline (0.025) & -0.077*\newline (0.026) & -0.069\newline (0.027) & -0.075*\newline (0.029) \\
Female authors & 0.020\newline (0.024) & -0.014\newline (0.025) & -0.023\newline (0.026) & -0.041\newline (0.028) \\
Native English & -0.065*\newline (0.025) & -0.026\newline (0.026) & 0.037\newline (0.027) & 0.002\newline (0.029) \\
\textit{Paper characteristics} &  &  &  \\
Formula density & -0.097*\newline (0.035) & -0.169***\newline (0.035) & -0.139***\newline (0.037) & -0.291***\newline (0.039) \\
Fig.\ \& Tab.\ density & -0.015\newline (0.033) & 0.057\newline (0.032) & 0.015\newline (0.033) & 0.102**\newline (0.036) \\
Table info density & 0.015\newline (0.031) & 0.006\newline (0.031) & -0.021\newline (0.032) & -0.029\newline (0.034) \\
Citation density & -0.141***\newline (0.038) & -0.023\newline (0.038) & -0.005\newline (0.039) & 0.001\newline (0.042) \\
Appendix ratio & -0.233***\newline (0.042) & -0.174***\newline (0.042) & -0.031\newline (0.043) & -0.174***\newline (0.046) \\
Log body length & -0.160***\newline (0.036) & -0.107**\newline (0.035) & -0.062\newline (0.036) & -0.069\newline (0.039) \\
Title length & 0.913***\newline (0.034) & 0.828***\newline (0.034) & 0.742***\newline (0.035) & 0.755***\newline (0.038) \\
Avg formula length & -0.046\newline (0.031) & -0.045\newline (0.032) & 0.003\newline (0.033) & -0.058\newline (0.035) \\
Avg sentence length & 0.055\newline (0.031) & 0.082*\newline (0.031) & 0.068\newline (0.032) & 0.119**\newline (0.034) \\
Abstract length & -0.030\newline (0.031) & -0.015\newline (0.031) & -0.048\newline (0.032) & -0.084*\newline (0.035) \\
\multicolumn{5}{P{\dimexpr\linewidth-2\tabcolsep-2\arrayrulewidth\relax}}{\textit{Notes.} Entries report ATO estimates from the primary specification.
Paper-clustered cross-fitted influence-function standard errors are in
parentheses. Nonbinary treatments contrast the upper and lower terciles.
Positive estimates indicate greater AI-human rating bias. Because treatment
contrasts and overlap populations differ across characteristics, each
row describes its own comparison. Benjamini-Hochberg
correction is applied separately within each AI model and characteristic
block. *** $p_{\mathrm{FDR}}<0.001$,
** $p_{\mathrm{FDR}}<0.01$,
* $p_{\mathrm{FDR}}<0.05$.} \\
\bottomrule
\end{longtable}
\endgroup

Appendix ratio and top-institution affiliation have negative estimates in all four configurations and are significant in three. Author count and log body length are negative in all four and significant in two. Industry involvement changes sign between Claude and GPT. Estimates for first-author student status, female-author share, table-information density, and average formula length are statistically indistinguishable from zero after FDR adjustment. These secondary comparisons describe the wider pattern of manuscript-group differences. The main recurring manuscript associations concern title length and formula density.

We repeated the analysis with a 512-token SciBERT proxy and with primary-area adjustment omitted. Title-length and formula-density contrasts retain their signs and FDR significance across all four configurations. Appendix~\ref{app:robustness} reports all characteristics under both specifications (Tables~\ref{tab:rdr_dml_ato_no_area} and \ref{tab:causal_512_proxy}) and diagnostics for the specification without area adjustment (Figure~\ref{fig:rdr_dml_diagnostics_no_area}). Significance varies for secondary characteristics. The two focal contrasts remain consistent across specifications.

\section{Discussion}
The findings answer the three research questions through different properties of score correspondence. Prompt comparisons show that closer scoring levels can accompany little change or a decrease in rank correlation. Domain comparisons identify differences that remain after aggregate calibration. Feature comparisons locate recurring signed differences among observed manuscript groups. Together, these results make the interpretation of an LLM score conditional on the scoring procedure, evaluation context, and manuscript information represented by the evidence.

\subsection{Scoring Rules and the Meaning of Agreement}
RQ1 shows why selecting a prompt requires specifying the evaluation property of interest. V4 reduces DeepSeek's mean divergence from 1.460 to 0.199 and MAE from 1.577 to 0.942. Rank correlation changes from 0.377 to 0.387. V5 brings mean divergence to 0.052 and yields a lower correlation of 0.321. The standard condition produces the highest observed correlation, 0.423, alongside larger absolute deviations. Different prompts therefore offer different combinations of correspondence on the same papers.

These comparisons extend the prompt-based evidence of \citet{drori2024human} and complement the procedural concerns described by \citet{thelwall2025complex}. V5 contains an explicit downward scoring instruction as well as historical references. Its proximity to the human mean demonstrates the behavior of that scoring rule. V4's lower MAE supplies additional paper-level evidence. A choice among these procedures should follow the intended interpretation of the scores, with each relevant property evaluated directly.

The prompt results give practical significance to the distinction between agreement and correlation \citep{bland1986agreement}. A prompt chosen for average alignment can differ from one chosen for ordering. Independent validation should therefore evaluate the properties required by the intended assessment task.

\subsection{Contextual Evidence After Aggregate Alignment}
RQ2 concerns whether aggregate correspondence describes the domains represented in the sample. Global affine calibration aligns overall location and spread by construction and preserves ranks. All four configurations nevertheless show negative mean divergence in Learning Theory. Societal AI Ethics combines positive divergence with weaker rank agreement. This pattern identifies contextual variation that a single aggregate adjustment leaves in place.

The result complements field differences in societal-impact evaluation \citep{kousha2025societal} and model-dependent language differences \citep{zhu2026language}. Our contribution is the joint examination of domain-level score differences, rank correspondence, and human consistency after aggregate alignment.

Domain ICC describes the reliability of the human mean used for comparison. This complements evidence of limited reviewer agreement \citep{bornmann2010reliability} and variation in review focus associated with reviewer confidence \citep{xie2026confidence}. Model-human correlations should therefore be interpreted alongside human consistency in the relevant domain.

\subsection{Manuscript Representation and Recurring Differences}
RQ3 identifies two recurring associations. Longer titles have positive adjusted contrasts in signed rating divergence across all four configurations. Formula density has negative contrasts. Both retain their directions and FDR significance under the alternative text proxy and area-adjustment specifications. They identify manuscript groups for which a single overall agreement summary is incomplete.

The findings connect document-feature research \citep{kousha2024factors} to differences between evaluation procedures. The outcome here is signed model-human divergence. This complements novelty prediction \citep{wu2025novelty} and controlled presentation studies \citep{du2025titletrap,yang2026presentation}. Our comparisons retain the content variation of real submissions, with recurring associations across models identifying manuscript groups for validation.

The contrasts condition on observed variables and text information associated with human ratings. Each feature defines a different comparison and overlap population. Author-identity effects in human review \citep{tomkins2017reviewer} also show why divergence can reflect differences arising in either evaluation procedure.

\subsection{Implications for Scholarly Assessment Information}
The Leiden Manifesto places quantitative evidence within expert-led, contextual research evaluation \citep{hicks2015leiden}. Its extension to LLMs emphasizes evaluation purpose and transparency \citep{thelwall2026responsible}. Our findings give these principles a concrete application: report signed differences and MAE for score levels and individual distances, rank evidence for ordering, and domain or manuscript-group results for the scope of correspondence. Describing the human reference makes these interpretations explicit.

The implications concern validation design. Studies of authorized assessment settings can test whether displaying model scores improves human judgments or decisions. Such studies should establish the added value of the score, including how people respond to disagreement. Similar error patterns across models also make error dependence relevant before their scores are combined. Operational use requires appropriate authorization, confidentiality protection, and human responsibility \citep{kankanhalli2024peer,shmueli2024editorial}.

\subsection{Limitations and Further Research}
The evidence concerns public ICLR 2024 submissions and four model configurations. Prompt comparisons use DeepSeek, and V4 and V5 change several instructions together. Standard-prompt selection and affine calibration use the analyzed sample. Independent years, venues, and model-by-prompt comparisons would establish the stability of the reported patterns. Model-generated confidence is a statement of certainty, whose calibration would require separate evaluation.

Human ratings contain noise and possible systematic bias. They define the recorded review-process reference in this study. Human-rating strata also condition on a component of the divergence outcome, introducing mathematical coupling and range restriction. The cell-level domain tests involve overlapping manuscript sets across models. These analyses provide descriptive and exploratory evidence alongside the domain-specific estimates.

Feature comparisons depend on observed adjustment variables, the generated human-rating proxy, and the corresponding overlap populations. The influence-function standard errors are conditional on that proxy and do not separately propagate uncertainty from its training. Dependence between the two learning stages also affects inference. Controlled manuscript changes and appropriately nested estimation would strengthen causal and inferential conclusions.

PDF extraction can affect the mathematical and tabular information supplied to a model. Public records may have been available during model development. Further work can document input coverage, use newly released or explicitly authorized manuscripts, and repeat scoring calls to quantify output variability. The consequences of using scores for editorial or research-assessment decisions require direct studies of those uses.

\section{Conclusion}
LLM-human score correspondence changes across prompts, domains, and manuscript groups. Area Chair guidance reduces average divergence and absolute error, while historical references with downward scoring instructions produce closer mean scores and lower rank correlation. Domain differences persist after global calibration. Title length and formula density have recurring adjusted associations with signed divergence across four model configurations. These findings connect the interpretation of scholarly assessment information to the procedure that generates it and the context in which it is compared. Evaluating LLM scores therefore requires evidence suited to the intended claim about score levels, manuscript ordering, or correspondence within manuscript groups.

\clearpage
\setstretch{1.15}
\printbibliography[heading=bibintoc,title={References}]
\clearpage
\par
\begingroup
\centering
{\Large\bfseries Appendices\par}
\vspace{8pt}
{\normalsize Interpreting LLM Scores in Peer Review: Alignment and Divergence Across Prompts, Domains, and Manuscript Features\par}
\endgroup
\vspace{12pt}
These appendices report the prompt specifications, extended results, human-reference comparisons, regression estimates, sensitivity analyses, and estimation details. Citations refer to the preceding reference list.
\setstretch{1.15}
\appendix
\setcounter{section}{0}
\setcounter{secnumdepth}{1}
\titleformat{\section}[display]{\centering\normalsize\bfseries}{Appendix~\thesection}{0pt}{}
\counterwithin*{table}{section}
\counterwithin*{figure}{section}
\setcounter{table}{0}
\setcounter{figure}{0}
\renewcommand{\thetable}{\thesection\arabic{table}}
\renewcommand{\thefigure}{\thesection\arabic{figure}}
\input{Appendices.tex}

\end{document}

%% file: Appendices.tex
\section{Prompt Design}\label{app:prompt_design}
The V1-V5 conditions combine structured review instructions with reviewer guidelines, ethics guidance, selectivity, and score adjustment. The tone and depth conditions use the same numerical scales in a separate review form. Table~\ref{tab:prompt_taxonomy} summarizes the ten conditions. Appendix~\ref{app:prompt_exact} specifies the scales, review forms, and distinctive instructions.

\begingroup
\small\setstretch{1.05}
\setlength{\tabcolsep}{3pt}
\renewcommand{\arraystretch}{1}
\setlength{\StudyTableWidth}{\dimexpr\linewidth-4\tabcolsep-3\arrayrulewidth\relax}
\begin{longtable}{P{0.2\StudyTableWidth}P{0.8\StudyTableWidth}}
\caption{Prompt Conditions Evaluated within DeepSeek-V3.2}\label{tab:prompt_taxonomy}\\
\toprule
\textbf{Condition} & \textbf{Instruction design} \\
\midrule
\endfirsthead
\multicolumn{2}{P{\dimexpr\linewidth-2\tabcolsep-2\arrayrulewidth\relax}}{\textit{Table~\thetable\ (continued)}} \\
\textbf{Condition} & \textbf{Instruction design} \\
\midrule
\endhead
\endfoot
\endlastfoot
V1 & Paper, ICLR review form, and baseline evaluation rubric. \\
V2 & V1 plus reviewer guidelines emphasizing evidence and methodological scrutiny. \\
V3 & V2 plus conduct and ethics guidance. \\
V4 & Area Chair selectivity added to the V3 framework. Choose the lower rating when undecided. \\
V5 & Conservative review instructions, historical score references, and explicit downward rating adjustment. \\
Constructive & Feedback-oriented guidance emphasizing actionable improvements. \\
Standard & Neutral guidance balancing strengths and weaknesses. \\
Critical & Skeptical guidance emphasizing assumptions, missing evidence, and overclaims. \\
Domain expert & Guidance emphasizing technical depth and field-specific positioning. \\
Quick & Time-constrained guidance focusing on decisive strengths and weaknesses. \\
\bottomrule
\end{longtable}
\endgroup

\section{Extended Prompt Comparisons}\label{app:prompt_results}
Tables~\ref{tab:appendix_prompt_soundness}-\ref{tab:appendix_prompt_confidence} report DeepSeek-V3.2 comparisons for Soundness, Presentation, Contribution, and Confidence. Table~\ref{tab:rating_integrated_sd} reports Rating and exact agreement.

\begin{landscape}
\begingroup
\small\setstretch{1.05}
\setlength{\tabcolsep}{3pt}
\renewcommand{\arraystretch}{1}
\setlength{\StudyTableWidth}{\dimexpr\linewidth-14\tabcolsep-8\arrayrulewidth\relax}
\begin{longtable}{P{0.17\StudyTableWidth}C{0.18\StudyTableWidth}C{0.18\StudyTableWidth}C{0.1\StudyTableWidth}C{0.1\StudyTableWidth}C{0.1\StudyTableWidth}C{0.17\StudyTableWidth}}
\caption{Prompt-Condition Performance on Soundness}\label{tab:appendix_prompt_soundness}\\
\toprule
\textbf{Prompt} & \textbf{Mean bias (SD)} & \textbf{MAE (SD)} & \textbf{Over (\%)} & \textbf{Under (\%)} & \textbf{Exact (\%)} & \textbf{Spearman $\rho$} \\
\midrule
\endfirsthead
\multicolumn{7}{P{\dimexpr\linewidth-2\tabcolsep-2\arrayrulewidth\relax}}{\textit{Table~\thetable\ (continued)}} \\
\textbf{Prompt} & \textbf{Mean bias (SD)} & \textbf{MAE (SD)} & \textbf{Over (\%)} & \textbf{Under (\%)} & \textbf{Exact (\%)} & \textbf{Spearman $\rho$} \\
\midrule
\endhead
\endfoot
\endlastfoot
V1 & 0.559 (0.421) & 0.592 (0.374) & 87.28 & 6.49 & 6.23 & 0.299 \\
V2 & 0.506 (0.436) & 0.557 (0.368) & 83.65 & 8.95 & 7.40 & 0.308 \\
V3 & 0.491 (0.432) & 0.545 (0.362) & 82.79 & 9.35 & 7.86 & 0.326 \\
V4 & 0.186 (0.482) & 0.409 (0.316) & 57.86 & 29.66 & 12.48 & 0.301 \\
V5 & 0.294 (0.474) & 0.447 (0.334) & 67.01 & 21.95 & 11.04 & 0.317 \\
constructive & 0.182 (0.471) & 0.403 (0.304) & 57.97 & 29.56 & 12.47 & 0.314 \\
domain expert & 0.272 (0.466) & 0.433 (0.322) & 65.54 & 22.79 & 11.68 & 0.344 \\
critical & -0.076 (0.430) & 0.348 (0.264) & 34.53 & 51.05 & 14.41 & 0.229 \\
quick & 0.278 (0.469) & 0.438 (0.325) & 66.23 & 22.39 & 11.38 & 0.331 \\
standard & 0.234 (0.464) & 0.414 (0.314) & 62.32 & 25.26 & 12.42 & 0.348 \\
\multicolumn{7}{P{\dimexpr\linewidth-2\tabcolsep-2\arrayrulewidth\relax}}{\textit{Note:} Entries use DeepSeek-V3.2 on the same 7,083 papers. Over and Under indicate scores above and below the human mean. Exact is the percentage of scores equal to that mean. Table~\ref{tab:omnibus_prompt_differences} reports omnibus comparisons.} \\
\bottomrule
\end{longtable}
\endgroup
\end{landscape}

\begin{landscape}
\begingroup
\small\setstretch{1.05}
\setlength{\tabcolsep}{3pt}
\renewcommand{\arraystretch}{1}
\setlength{\StudyTableWidth}{\dimexpr\linewidth-14\tabcolsep-8\arrayrulewidth\relax}
\begin{longtable}{P{0.17\StudyTableWidth}C{0.18\StudyTableWidth}C{0.18\StudyTableWidth}C{0.1\StudyTableWidth}C{0.1\StudyTableWidth}C{0.1\StudyTableWidth}C{0.17\StudyTableWidth}}
\caption{Prompt-Condition Performance on Presentation}\label{tab:appendix_prompt_presentation}\\
\toprule
\textbf{Prompt} & \textbf{Mean bias (SD)} & \textbf{MAE (SD)} & \textbf{Over (\%)} & \textbf{Under (\%)} & \textbf{Exact (\%)} & \textbf{Spearman $\rho$} \\
\midrule
\endfirsthead
\multicolumn{7}{P{\dimexpr\linewidth-2\tabcolsep-2\arrayrulewidth\relax}}{\textit{Table~\thetable\ (continued)}} \\
\textbf{Prompt} & \textbf{Mean bias (SD)} & \textbf{MAE (SD)} & \textbf{Over (\%)} & \textbf{Under (\%)} & \textbf{Exact (\%)} & \textbf{Spearman $\rho$} \\
\midrule
\endhead
\endfoot
\endlastfoot
V1 & 0.422 (0.477) & 0.519 (0.369) & 77.26 & 14.27 & 8.47 & 0.333 \\
V2 & 0.355 (0.487) & 0.487 (0.355) & 71.95 & 18.59 & 9.46 & 0.344 \\
V3 & 0.353 (0.488) & 0.486 (0.356) & 71.74 & 18.52 & 9.74 & 0.336 \\
V4 & 0.187 (0.515) & 0.439 (0.328) & 58.00 & 31.13 & 10.87 & 0.301 \\
V5 & 0.234 (0.513) & 0.452 (0.338) & 61.50 & 27.87 & 10.63 & 0.288 \\
constructive & 0.165 (0.492) & 0.415 (0.311) & 56.01 & 32.32 & 11.68 & 0.343 \\
domain expert & 0.150 (0.488) & 0.408 (0.307) & 54.98 & 33.16 & 11.86 & 0.360 \\
critical & 0.088 (0.490) & 0.397 (0.300) & 49.30 & 38.30 & 12.40 & 0.340 \\
quick & 0.169 (0.490) & 0.415 (0.310) & 56.97 & 31.41 & 11.62 & 0.353 \\
standard & 0.204 (0.481) & 0.420 (0.311) & 60.27 & 28.62 & 11.11 & 0.376 \\
\multicolumn{7}{P{\dimexpr\linewidth-2\tabcolsep-2\arrayrulewidth\relax}}{\textit{Note:} Entries use DeepSeek-V3.2 on the same 7,083 papers. Over and Under indicate scores above and below the human mean. Exact is the percentage of scores equal to that mean. Table~\ref{tab:omnibus_prompt_differences} reports omnibus comparisons.} \\
\bottomrule
\end{longtable}
\endgroup
\end{landscape}

\begin{landscape}
\begingroup
\small\setstretch{1.05}
\setlength{\tabcolsep}{3pt}
\renewcommand{\arraystretch}{1}
\setlength{\StudyTableWidth}{\dimexpr\linewidth-14\tabcolsep-8\arrayrulewidth\relax}
\begin{longtable}{P{0.17\StudyTableWidth}C{0.18\StudyTableWidth}C{0.18\StudyTableWidth}C{0.1\StudyTableWidth}C{0.1\StudyTableWidth}C{0.1\StudyTableWidth}C{0.17\StudyTableWidth}}
\caption{Prompt-Condition Performance on Contribution}\label{tab:appendix_prompt_contribution}\\
\toprule
\textbf{Prompt} & \textbf{Mean bias (SD)} & \textbf{MAE (SD)} & \textbf{Over (\%)} & \textbf{Under (\%)} & \textbf{Exact (\%)} & \textbf{Spearman $\rho$} \\
\midrule
\endfirsthead
\multicolumn{7}{P{\dimexpr\linewidth-2\tabcolsep-2\arrayrulewidth\relax}}{\textit{Table~\thetable\ (continued)}} \\
\textbf{Prompt} & \textbf{Mean bias (SD)} & \textbf{MAE (SD)} & \textbf{Over (\%)} & \textbf{Under (\%)} & \textbf{Exact (\%)} & \textbf{Spearman $\rho$} \\
\midrule
\endhead
\endfoot
\endlastfoot
V1 & 0.691 (0.477) & 0.728 (0.418) & 89.64 & 5.89 & 4.48 & 0.301 \\
V2 & 0.680 (0.481) & 0.720 (0.420) & 89.19 & 6.18 & 4.63 & 0.300 \\
V3 & 0.699 (0.470) & 0.733 (0.415) & 90.06 & 5.58 & 4.36 & 0.309 \\
V4 & 0.232 (0.486) & 0.428 (0.327) & 62.26 & 26.32 & 11.42 & 0.276 \\
V5 & 0.201 (0.470) & 0.405 (0.312) & 59.90 & 27.93 & 12.17 & 0.254 \\
constructive & 0.704 (0.458) & 0.732 (0.411) & 91.01 & 4.79 & 4.21 & 0.327 \\
domain expert & 0.697 (0.458) & 0.727 (0.409) & 90.77 & 5.00 & 4.24 & 0.345 \\
critical & 0.511 (0.488) & 0.588 (0.393) & 81.12 & 11.63 & 7.24 & 0.335 \\
quick & 0.718 (0.450) & 0.741 (0.411) & 91.99 & 3.92 & 4.08 & 0.343 \\
standard & 0.723 (0.453) & 0.747 (0.412) & 91.97 & 4.25 & 3.78 & 0.327 \\
\multicolumn{7}{P{\dimexpr\linewidth-2\tabcolsep-2\arrayrulewidth\relax}}{\textit{Note:} Entries use DeepSeek-V3.2 on the same 7,083 papers. Over and Under indicate scores above and below the human mean. Exact is the percentage of scores equal to that mean. Table~\ref{tab:omnibus_prompt_differences} reports omnibus comparisons.} \\
\bottomrule
\end{longtable}
\endgroup
\end{landscape}

\begin{landscape}
\begingroup
\small\setstretch{1.05}
\setlength{\tabcolsep}{3pt}
\renewcommand{\arraystretch}{1}
\setlength{\StudyTableWidth}{\dimexpr\linewidth-14\tabcolsep-8\arrayrulewidth\relax}
\begin{longtable}{P{0.17\StudyTableWidth}C{0.18\StudyTableWidth}C{0.18\StudyTableWidth}C{0.1\StudyTableWidth}C{0.1\StudyTableWidth}C{0.1\StudyTableWidth}C{0.17\StudyTableWidth}}
\caption{Prompt-Condition Performance on Confidence}\label{tab:appendix_prompt_confidence}\\
\toprule
\textbf{Prompt} & \textbf{Mean bias (SD)} & \textbf{MAE (SD)} & \textbf{Over (\%)} & \textbf{Under (\%)} & \textbf{Exact (\%)} & \textbf{Spearman $\rho$} \\
\midrule
\endfirsthead
\multicolumn{7}{P{\dimexpr\linewidth-2\tabcolsep-2\arrayrulewidth\relax}}{\textit{Table~\thetable\ (continued)}} \\
\textbf{Prompt} & \textbf{Mean bias (SD)} & \textbf{MAE (SD)} & \textbf{Over (\%)} & \textbf{Under (\%)} & \textbf{Exact (\%)} & \textbf{Spearman $\rho$} \\
\midrule
\endhead
\endfoot
\endlastfoot
V1 & 0.358 (0.490) & 0.477 (0.375) & 67.63 & 15.43 & 16.94 & 0.008 \\
V2 & 0.358 (0.491) & 0.478 (0.375) & 67.68 & 15.45 & 16.87 & 0.006 \\
V3 & 0.358 (0.490) & 0.477 (0.375) & 67.61 & 15.42 & 16.97 & 0.020 \\
V4 & 0.357 (0.489) & 0.476 (0.374) & 67.56 & 15.45 & 17.00 & -0.007 \\
V5 & 0.356 (0.488) & 0.475 (0.373) & 67.51 & 15.46 & 17.03 & 0.013 \\
constructive & 0.344 (0.499) & 0.476 (0.374) & 66.81 & 16.18 & 17.01 & 0.016 \\
domain expert & 0.357 (0.488) & 0.476 (0.373) & 67.60 & 15.42 & 16.98 & 0.027 \\
critical & 0.344 (0.494) & 0.474 (0.372) & 66.86 & 16.07 & 17.07 & 0.046 \\
quick & 0.342 (0.498) & 0.475 (0.373) & 66.65 & 16.26 & 17.08 & 0.033 \\
standard & 0.331 (0.508) & 0.476 (0.376) & 66.03 & 16.86 & 17.11 & 0.030 \\
\multicolumn{7}{P{\dimexpr\linewidth-2\tabcolsep-2\arrayrulewidth\relax}}{\textit{Note:} Entries use DeepSeek-V3.2 on the same 7,083 papers. Over and Under indicate scores above and below the human mean. Exact is the percentage of scores equal to that mean. Table~\ref{tab:omnibus_prompt_differences} reports omnibus comparisons.} \\
\bottomrule
\end{longtable}
\endgroup
\end{landscape}

\begin{landscape}
\begingroup
\small\setstretch{1.05}
\setlength{\tabcolsep}{3pt}
\renewcommand{\arraystretch}{1}
\setlength{\StudyTableWidth}{\dimexpr\linewidth-14\tabcolsep-8\arrayrulewidth\relax}
\begin{longtable}{P{0.17\StudyTableWidth}C{0.14\StudyTableWidth}C{0.14\StudyTableWidth}C{0.14\StudyTableWidth}C{0.14\StudyTableWidth}C{0.13\StudyTableWidth}C{0.14\StudyTableWidth}}
\caption{Omnibus Tests for Prompt-Condition Differences Across Dimensions}\label{tab:omnibus_prompt_differences}\\
\toprule
\textbf{Dimension} & \textbf{Bias\newline Friedman $\chi^2$} & \textbf{MAE\newline Friedman $\chi^2$} & \textbf{Over\newline Cochran's $Q$} & \textbf{Under\newline Cochran's $Q$} & \textbf{Exact\newline Cochran's $Q$} & \textbf{Permutation $\rho$} \\
\midrule
\endfirsthead
\multicolumn{7}{P{\dimexpr\linewidth-2\tabcolsep-2\arrayrulewidth\relax}}{\textit{Table~\thetable\ (continued)}} \\
\textbf{Dimension} & \textbf{Bias\newline Friedman $\chi^2$} & \textbf{MAE\newline Friedman $\chi^2$} & \textbf{Over\newline Cochran's $Q$} & \textbf{Under\newline Cochran's $Q$} & \textbf{Exact\newline Cochran's $Q$} & \textbf{Permutation $\rho$} \\
\midrule
\endhead
\endfoot
\endlastfoot
Soundness & $2.379\times10^{4}{}^{***}$ & $5004^{***}$ & $1.481\times10^{4}{}^{***}$ & $1.257\times10^{4}{}^{***}$ & $1297^{***}$ & $0.1189^{**}$ \\
Presentation & $9451^{***}$ & $1667^{***}$ & $5895^{***}$ & $4996^{***}$ & $377.2^{***}$ & $0.0877^{**}$ \\
Contribution & $2.700\times10^{4}{}^{***}$ & $1.393\times10^{4}{}^{***}$ & $1.320\times10^{4}{}^{***}$ & $9976^{***}$ & $2319^{***}$ & $0.0913^{**}$ \\
Rating & $3.931\times10^{4}{}^{***}$ & $1.061\times10^{4}{}^{***}$ & $1.619\times10^{4}{}^{***}$ & $1.494\times10^{4}{}^{***}$ & $483.5^{***}$ & $0.1021^{**}$ \\
Confidence & $847.7^{***}$ & $19.30^{*}$ & $506.2^{***}$ & $458.3^{***}$ & $14.87$ & $0.0530^{*}$ \\
\multicolumn{7}{P{\dimexpr\linewidth-2\tabcolsep-2\arrayrulewidth\relax}}{\textit{Notes:} \(N=7{,}083\). Bias and MAE use Friedman \(\chi^2\) tests across the ten conditions. Over, Under, and Exact use Cochran's \(Q\) tests. The permutation column reports the observed range of Spearman \(\rho\), with significance assessed by permutation. Stars apply to these omnibus tests: \(^{***}p<0.001\), \(^{**}p<0.01\), \(^{*}p<0.05\). Exact agreement compares an integer LLM score with the mean of three to five human scores, so it also reflects the numerical form of the reference.} \\
\bottomrule
\end{longtable}
\endgroup
\end{landscape}

\clearpage\section{Human Reference and Contextual Heterogeneity}\label{app:context}
The area-model cell decomposition is reported as exploratory. Model-specific cells within an area use overlapping manuscripts. The descriptive domain means and correlations are interpreted separately from the cell-level significance tests.

\begin{figure}[htbp]
\centering
\includegraphics[width=0.9\linewidth]{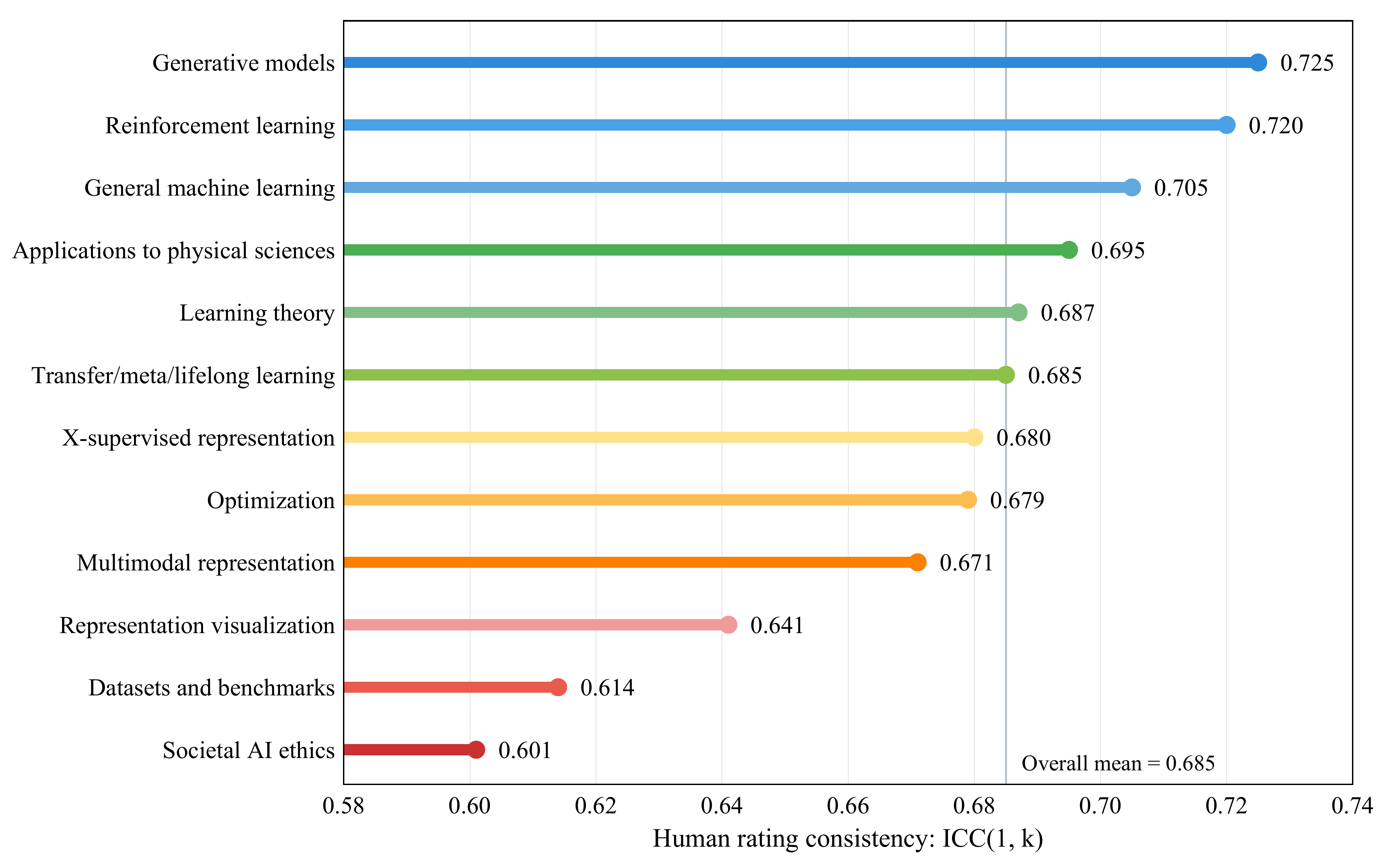}
\caption{Human Rating Consistency Across Domains. The figure covers the 12 primary areas with at least 200 papers. ICC(1,$k$) measures consistency of average human ratings. Differences in human consensus help interpret model-human rank agreement.}
\label{fig:area_human_icc}
\end{figure}

Figure~\ref{fig:area_human_icc} reports human rating consistency in the 12 areas with at least 200 papers. Table~\ref{tab:kruskal_wallis_all_dims} compares absolute disagreement among human reviewers across these areas. Table~\ref{tab:anova_rating_full} decomposes model and area variation in 28 cells covering 3,684 papers and 14,736 model-paper scores. Its partial effect sizes summarize weighted cell variation. Tables~\ref{tab:rating_bin_variance_summary} and \ref{tab:rating_bin_full} report dispersion and heterogeneity within the three rating bins. Table~\ref{tab:area_struct_stats_compact_sd} describes manuscript structure in the seven displayed areas. Its appendix/main-text ratio uses a different denominator from the appendix/total ratio in the manuscript-level regressions.

\begingroup
\small\setstretch{1.05}
\setlength{\tabcolsep}{3pt}
\renewcommand{\arraystretch}{1}
\setlength{\StudyTableWidth}{\dimexpr\linewidth-8\tabcolsep-5\arrayrulewidth\relax}
\begin{longtable}{P{0.35\StudyTableWidth}C{0.3\StudyTableWidth}C{0.1\StudyTableWidth}C{0.25\StudyTableWidth}}
\caption{Kruskal-Wallis Tests of Domain Differences in Human Reviewer Disagreement}\label{tab:kruskal_wallis_all_dims}\\
\toprule
\textbf{Dimension} & \textbf{Test statistic $H$} & \textbf{df} & \textbf{$p_{\mathrm{FDR}}$} \\
\midrule
\endfirsthead
\multicolumn{4}{P{\dimexpr\linewidth-2\tabcolsep-2\arrayrulewidth\relax}}{\textit{Table~\thetable\ (continued)}} \\
\textbf{Dimension} & \textbf{Test statistic $H$} & \textbf{df} & \textbf{$p_{\mathrm{FDR}}$} \\
\midrule
\endhead
\endfoot
\endlastfoot
Soundness & 22.49 & 11 & 0.026 \\
Presentation & 44.67 & 11 & \(<0.001\) \\
Contribution & 40.51 & 11 & \(<0.001\) \\
Rating & 28.14 & 11 & 0.005 \\
Confidence & 20.39 & 11 & 0.040 \\
\multicolumn{4}{P{\dimexpr\linewidth-2\tabcolsep-2\arrayrulewidth\relax}}{\textit{Note:} Human reviewer disagreement is measured at the paper level as
the mean pairwise absolute difference among valid human-review scores,
\(D^{HH}_{i,d}\). Tests compare the 12 primary research areas with at least
200 papers. \(p\)-values are adjusted across the five dimensions using the
Benjamini-Hochberg procedure.} \\
\bottomrule
\end{longtable}
\endgroup

\begin{landscape}
\begingroup
\small\setstretch{1.05}
\setlength{\tabcolsep}{3pt}
\renewcommand{\arraystretch}{1}
\setlength{\StudyTableWidth}{\dimexpr\linewidth-16\tabcolsep-9\arrayrulewidth\relax}
\begin{longtable}{P{0.145\StudyTableWidth}C{0.13\StudyTableWidth}C{0.17\StudyTableWidth}C{0.095\StudyTableWidth}C{0.1\StudyTableWidth}C{0.17\StudyTableWidth}C{0.095\StudyTableWidth}C{0.095\StudyTableWidth}}
\caption{Full Wild-Bootstrap Heterogeneity Decomposition for \textit{Rating}}\label{tab:anova_rating_full}\\
\toprule
\textbf{Metric} & \textbf{Effect} & \textbf{Raw $F(df_1,df_2)$} & \textbf{Raw $q_{\mathrm{FDR}}$} & \textbf{Raw partial $\eta^2$} & \textbf{Calibrated $F(df_1,df_2)$} & \textbf{Calibrated $q_{\mathrm{FDR}}$} & \textbf{Calibrated partial $\eta^2$} \\
\midrule
\endfirsthead
\multicolumn{8}{P{\dimexpr\linewidth-2\tabcolsep-2\arrayrulewidth\relax}}{\textit{Table~\thetable\ (continued)}} \\
\textbf{Metric} & \textbf{Effect} & \textbf{Raw $F(df_1,df_2)$} & \textbf{Raw $q_{\mathrm{FDR}}$} & \textbf{Raw partial $\eta^2$} & \textbf{Calibrated $F(df_1,df_2)$} & \textbf{Calibrated $q_{\mathrm{FDR}}$} & \textbf{Calibrated partial $\eta^2$} \\
\midrule
\endhead
\endfoot
\endlastfoot
Mean Bias & Area & 40.94 (6,18) & 0.0003 & 0.932 & 20.85 (6,18) & 0.0030 & 0.874 \\
Mean Bias & Model & 205.5 (3,18) & 0.0003 & 0.972 & 0.000 (3,18) & 1.000 & 0.000 \\
Mean Bias & Non-additivity & 0.147 (1,17) & 0.7840 & 0.009 & --- & --- & --- \\
MAE & Area & 31.68 (6,18) & 0.0006 & 0.913 & 0.672 (6,18) & 0.5680 & 0.183 \\
MAE & Model & 173.4 (3,18) & 0.0006 & 0.967 & 28.64 (3,18) & 0.0018 & 0.827 \\
MAE & Non-additivity & 0.393 (1,17) & 0.4820 & 0.023 & 1.852 (1,17) & 0.5680 & 0.098 \\
Spearman $\rho$ & Area & 25.40 (6,18) & 0.0003 & 0.894 & 25.40 (6,18) & 0.0003 & 0.894 \\
Spearman $\rho$ & Model & 210.7 (3,18) & 0.0003 & 0.972 & 210.7 (3,18) & 0.0003 & 0.972 \\
Spearman $\rho$ & Non-additivity & 0.236 (1,17) & 0.7600 & 0.014 & 0.236 (1,17) & 0.7530 & 0.014 \\
\multicolumn{8}{P{\dimexpr\linewidth-2\tabcolsep-2\arrayrulewidth\relax}}{\textit{Notes:} Entries report wild-bootstrap nested-$F$ tests and corresponding partial \(\eta^2\) effect sizes. Non-additivity is tested using Tukey's 1-df procedure. \(q_{\mathrm{FDR}}\) denotes BH-FDR-adjusted \(p\)-values within each metric \(\times\) specification block. \(N=14{,}736\) and \(n_{\mathrm{cells}}=28\) for all rows. `---' indicates not applicable.} \\
\bottomrule
\end{longtable}
\endgroup
\end{landscape}

\begingroup
\small\setstretch{1.05}
\setlength{\tabcolsep}{3pt}
\renewcommand{\arraystretch}{1}
\setlength{\StudyTableWidth}{\dimexpr\linewidth-8\tabcolsep-5\arrayrulewidth\relax}
\begin{longtable}{P{0.23\StudyTableWidth}C{0.15\StudyTableWidth}C{0.31\StudyTableWidth}C{0.31\StudyTableWidth}}
\caption{Dispersion and Variance Heterogeneity Across Rating Bins}\label{tab:rating_bin_variance_summary}\\
\toprule
\multicolumn{4}{P{\dimexpr\linewidth-2\tabcolsep-2\arrayrulewidth\relax}}{\textbf{Panel A. Descriptive dispersion statistics}} \\
\textbf{Bin} & \textbf{\(N\)} & \textbf{SD [95\% CI]} & \textbf{Variance [95\% CI]} \\
\midrule
\endfirsthead
\multicolumn{4}{P{\dimexpr\linewidth-2\tabcolsep-2\arrayrulewidth\relax}}{\textit{Table~\thetable\ (continued)}} \\
\multicolumn{4}{P{\dimexpr\linewidth-2\tabcolsep-2\arrayrulewidth\relax}}{\textbf{Panel A. Descriptive dispersion statistics}} \\
\textbf{Bin} & \textbf{\(N\)} & \textbf{SD [95\% CI]} & \textbf{Variance [95\% CI]} \\
\midrule
\endhead
\endfoot
\endlastfoot
Low & 2126 & 0.603 [0.578, 0.627] & 0.364 [0.334, 0.394] \\
Middle & 3764 & 0.555 [0.546, 0.562] & 0.308 [0.298, 0.316] \\
High & 1193 & 0.469 [0.448, 0.490] & 0.220 [0.201, 0.240] \\
\multicolumn{4}{P{\dimexpr\linewidth-2\tabcolsep-2\arrayrulewidth\relax}}{\textbf{Panel B. Variance heterogeneity}} \\
\multicolumn{2}{l}{\textbf{Comparison}} & \textbf{Statistic} & \textbf{\(p\)} \\
\multicolumn{2}{l}{Overall} & 51.25 & \(<0.001\) \\
\multicolumn{2}{l}{Low vs.\ Middle} & 5.948 & 0.015 \\
\multicolumn{2}{l}{Low vs.\ High} & 42.67 & \(<0.001\) \\
\multicolumn{2}{l}{Middle vs.\ High} & 135.5 & \(<0.001\) \\
\multicolumn{4}{P{\dimexpr\linewidth-2\tabcolsep-2\arrayrulewidth\relax}}{\textit{Notes:} Bins are based on half-up rounded human mean ratings. Confidence intervals are bootstrap percentile intervals. The overall result is based on Brown-Forsythe, and pairwise \(p\)-values are Holm-adjusted.} \\
\bottomrule
\end{longtable}
\endgroup

\begin{landscape}
\begingroup
\small\setstretch{1.05}
\setlength{\tabcolsep}{3pt}
\renewcommand{\arraystretch}{1}
\setlength{\StudyTableWidth}{\dimexpr\linewidth-20\tabcolsep-11\arrayrulewidth\relax}
\begin{longtable}{P{0.12\StudyTableWidth}C{0.1\StudyTableWidth}C{0.1\StudyTableWidth}C{0.08\StudyTableWidth}C{0.1\StudyTableWidth}C{0.1\StudyTableWidth}C{0.08\StudyTableWidth}C{0.12\StudyTableWidth}C{0.1\StudyTableWidth}C{0.1\StudyTableWidth}}
\caption{Heterogeneity Across Three Rating Bins on the \textit{Rating} Dimension}\label{tab:rating_bin_full}\\
\toprule
\textbf{Model} & \textbf{ANOVA $F$} & \textbf{$p$ (FDR)} & \textbf{$\eta^2$} & \textbf{Kruskal $H$} & \textbf{$p$ (FDR)} & \textbf{$\epsilon^2$} & \textbf{Cochran's $Q$} & \textbf{$p$ (FDR)} & \textbf{$I^2$} \\
\midrule
\endfirsthead
\multicolumn{10}{P{\dimexpr\linewidth-2\tabcolsep-2\arrayrulewidth\relax}}{\textit{Table~\thetable\ (continued)}} \\
\textbf{Model} & \textbf{ANOVA $F$} & \textbf{$p$ (FDR)} & \textbf{$\eta^2$} & \textbf{Kruskal $H$} & \textbf{$p$ (FDR)} & \textbf{$\epsilon^2$} & \textbf{Cochran's $Q$} & \textbf{$p$ (FDR)} & \textbf{$I^2$} \\
\midrule
\endhead
\endfoot
\endlastfoot
DeepSeek & 5373 & \(<0.001\) & 0.603 & 4280 & \(<0.001\) & 0.604 & 57.24 & \(<0.001\) & 96.51 \\
Claude & \(1.269\times10^4\) & \(<0.001\) & 0.782 & 5590 & \(<0.001\) & 0.789 & 61.90 & \(<0.001\) & 96.77 \\
Gemini & 2562 & \(<0.001\) & 0.420 & 2999 & \(<0.001\) & 0.423 & 62.23 & \(<0.001\) & 96.79 \\
GPT & 2359 & \(<0.001\) & 0.400 & 2875 & \(<0.001\) & 0.406 & 70.22 & \(<0.001\) & 97.15 \\
\multicolumn{10}{P{\dimexpr\linewidth-2\tabcolsep-2\arrayrulewidth\relax}}{\textit{Notes:} All analyses are based on the same three rating bins. \(\eta^2\), \(\epsilon^2\), and \(I^2\) report heterogeneity effect sizes. All \(p\)-values are FDR-adjusted.} \\
\bottomrule
\end{longtable}
\endgroup
\end{landscape}

\begin{landscape}
\begingroup
\small\setstretch{1.05}
\setlength{\tabcolsep}{3pt}
\renewcommand{\arraystretch}{1}
\setlength{\StudyTableWidth}{\dimexpr\linewidth-14\tabcolsep-8\arrayrulewidth\relax}
\begin{longtable}{P{0.19\StudyTableWidth}C{0.055\StudyTableWidth}C{0.145\StudyTableWidth}C{0.145\StudyTableWidth}C{0.145\StudyTableWidth}C{0.145\StudyTableWidth}C{0.175\StudyTableWidth}}
\caption{Structural Statistics by Primary Area}\label{tab:area_struct_stats_compact_sd}\\
\toprule
\textbf{Main area} & \textbf{$N$} & \textbf{Formulas} & \textbf{Figures} & \textbf{Tables} & \textbf{Document length (k)} & \textbf{Appendix/main-text ratio} \\
\midrule
\endfirsthead
\multicolumn{7}{P{\dimexpr\linewidth-2\tabcolsep-2\arrayrulewidth\relax}}{\textit{Table~\thetable\ (continued)}} \\
\textbf{Main area} & \textbf{$N$} & \textbf{Formulas} & \textbf{Figures} & \textbf{Tables} & \textbf{Document length (k)} & \textbf{Appendix/main-text ratio} \\
\midrule
\endhead
\endfoot
\endlastfoot
Societal AI ethics & 521 & 11.13 (8.45)\newline 24.10 (31.74) & 4.94 (3.30)\newline 7.92 (5.61) & 3.43 (2.35)\newline 6.87 (5.02) & 68.9 (8.2)\newline 162.2 (53.5) & 1.354 \\
Multi-modal rep. learn. & 1010 & 8.72 (6.27)\newline 12.90 (14.46) & 5.02 (3.15)\newline 7.60 (5.06) & 4.24 (2.36)\newline 7.17 (4.46) & 66.2 (8.2)\newline $\underline{\mathbf{153.3\ (48.1)}}$ & $\underline{\mathbf{1.315}}$ \\
Generative models & 758 & 10.38 (8.48)\newline 19.00 (24.44) & 5.63 (3.25)\newline $\overline{\mathbf{9.27\ (6.22)}}$ & 3.27 (2.17)\newline 5.90 (4.61) & 64.7 (8.7)\newline 154.2 (50.8) & 1.382 \\
Learning theory & 200 & 26.59 (11.66)\newline $\overline{\mathbf{107.9\ (88.36)}}$ & 3.12 (3.22)\newline $\underline{\mathbf{5.39\ (5.49)}}$ & 1.41 (1.61)\newline $\underline{\mathbf{3.23\ (3.84)}}$ & 70.5 (9.4)\newline 186.4 (75.1) & 1.645 \\
Reinforcement learning & 636 & 15.49 (10.80)\newline 38.11 (49.88) & 5.08 (3.39)\newline 8.27 (6.04) & 1.87 (1.60)\newline 4.43 (4.37) & 68.3 (9.0)\newline 159.9 (56.6) & 1.340 \\
Optimization & 297 & 22.35 (11.91)\newline 72.72 (77.77) & 4.15 (3.23)\newline 6.44 (5.28) & 2.10 (1.83)\newline 3.74 (3.40) & 67.1 (7.9)\newline 161.1 (60.3) & 1.402 \\
Datasets and benchmarks & 262 & 3.06 (3.93)\newline $\underline{\mathbf{5.87\ (10.34)}}$ & 5.04 (3.01)\newline 8.98 (6.90) & 3.88 (2.46)\newline $\overline{\mathbf{8.35\ (6.44)}}$ & 67.3 (7.7)\newline $\overline{\mathbf{191.6\ (75.4)}}$ & $\overline{\mathbf{1.846}}$ \\*
\bottomrule
\multicolumn{7}{P{\dimexpr\linewidth-2\tabcolsep-2\arrayrulewidth\relax}}{\textit{Note:} Each cell reports the mean and standard deviation for body-text statistics on the first line and full-document statistics on the second line. For each metric, the maximum and minimum full-mean values across areas are highlighted with boldface and overline/underline, respectively.} \\*
\bottomrule
\end{longtable}
\endgroup
\end{landscape}

\clearpage\section{Complete OLS Associations}\label{app:ols}
Table~\ref{tab:global_delta_ols_coef} reports the ex-ante OLS specification. Table~\ref{tab:global_delta_ols_filled_coef_B} adds human-rating dispersion and reviewer count. Both use 7,083 papers per model, and their $R^2$ values differ by at most 0.002. OLS coefficients describe changes per unit of each variable. DML estimates compare high and low groups within a characteristic-specific overlap population. These different scales and populations govern the interpretation of each estimate. Standard errors are HC3 robust, and OLS significance marks use unadjusted $p$-values.

\begin{landscape}
\begingroup
\small\setstretch{1.05}
\setlength{\tabcolsep}{3pt}
\renewcommand{\arraystretch}{1}
\setlength{\StudyTableWidth}{\dimexpr\linewidth-10\tabcolsep-6\arrayrulewidth\relax}
\begin{longtable}{P{0.26\StudyTableWidth}C{0.185\StudyTableWidth}C{0.185\StudyTableWidth}C{0.185\StudyTableWidth}C{0.185\StudyTableWidth}}
\caption{Multivariate Regression Estimates of AI-Human Rating Bias}\label{tab:global_delta_ols_coef}\\
\toprule
\textbf{Variable} & \textbf{Claude\newline Coef. (SE)} & \textbf{DeepSeek\newline Coef. (SE)} & \textbf{Gemini\newline Coef. (SE)} & \textbf{GPT\newline Coef. (SE)} \\
\midrule
\endfirsthead
\multicolumn{5}{P{\dimexpr\linewidth-2\tabcolsep-2\arrayrulewidth\relax}}{\textit{Table~\thetable\ (continued)}} \\
\textbf{Variable} & \textbf{Claude\newline Coef. (SE)} & \textbf{DeepSeek\newline Coef. (SE)} & \textbf{Gemini\newline Coef. (SE)} & \textbf{GPT\newline Coef. (SE)} \\
\midrule
\endhead
\endfoot
\endlastfoot
Intercept & 19.02 (1.362)*** & 9.467 (1.256)*** & 0.344 (1.346) & 2.478 (1.354) \\
\multicolumn{5}{P{\dimexpr\linewidth-2\tabcolsep-2\arrayrulewidth\relax}}{\textit{Author characteristics}} \\
Industry involvement & -0.137 (0.039)*** & -0.020 (0.039) & 0.046 (0.041) & 0.125 (0.043)** \\
Top institution & -0.196 (0.042)*** & -0.127 (0.043)** & -0.084 (0.044) & -0.059 (0.048) \\
Female authors & 0.037 (0.059) & -0.041 (0.059) & -0.091 (0.061) & -0.096 (0.067) \\
First author student & 0.018 (0.027) & 0.016 (0.027) & 0.007 (0.028) & -0.002 (0.030) \\
Native English & -0.134 (0.039)*** & -0.028 (0.038) & 0.101 (0.039)* & 0.043 (0.042) \\
N authors & -0.033 (0.005)*** & -0.019 (0.005)*** & -0.011 (0.005)* & -0.019 (0.006)*** \\
\multicolumn{5}{P{\dimexpr\linewidth-2\tabcolsep-2\arrayrulewidth\relax}}{\textit{Paper characteristics}} \\
Formula density & -0.742 (0.100)*** & -0.711 (0.097)*** & -0.368 (0.103)*** & -0.847 (0.109)*** \\
Fig. \& Tab. density & -1.033 (0.245)*** & -0.417 (0.240) & -0.120 (0.254) & -0.079 (0.260) \\
Table info density & 0.109 (0.174) & 0.146 (0.172) & -0.067 (0.175) & -0.177 (0.189) \\
Citation density & -0.082 (0.024)*** & 0.050 (0.024)* & 0.020 (0.026) & 0.110 (0.028)*** \\
Appendix ratio & -2.329 (0.135)*** & -1.751 (0.131)*** & -0.394 (0.145)** & -1.201 (0.148)*** \\
Log body length & -1.506 (0.120)*** & -0.740 (0.112)*** & 0.067 (0.120) & -0.119 (0.121) \\
Title length & 0.014 (0.0004)*** & 0.011 (0.0004)*** & 0.009 (0.0004)*** & 0.009 (0.0004)*** \\
Avg formula length & 9.9e-5 (1.2e-4) & 1.2e-4 (1.2e-4) & -1.4e-4 (1.1e-4) & -3.7e-5 (1.3e-4) \\
Avg sentence length & 0.0009 (0.0008) & 0.0010 (0.0008) & 0.0013 (0.0008) & 0.0015 (0.0008) \\
Abstract length & -8.2e-5 (2.2e-5)*** & -5.3e-5 (2.1e-5)* & -7.8e-5 (2.2e-5)*** & -8.6e-5 (2.4e-5)*** \\
Observations & 7,083 & 7,083 & 7,083 & 7,083 \\
$R^2$ & 0.339 & 0.209 & 0.092 & 0.106 \\
Adj. $R^2$ & 0.338 & 0.208 & 0.089 & 0.104 \\
\multicolumn{5}{P{\dimexpr\linewidth-2\tabcolsep-2\arrayrulewidth\relax}}{\textit{Note:} Standard errors are HC3 robust standard errors. *** $p<0.001$, ** $p<0.01$, * $p<0.05$.} \\
\bottomrule
\end{longtable}
\endgroup
\end{landscape}

\begin{landscape}
\begingroup
\small\setstretch{1.05}
\setlength{\tabcolsep}{3pt}
\renewcommand{\arraystretch}{1}
\setlength{\StudyTableWidth}{\dimexpr\linewidth-10\tabcolsep-6\arrayrulewidth\relax}
\begin{longtable}{P{0.26\StudyTableWidth}C{0.185\StudyTableWidth}C{0.185\StudyTableWidth}C{0.185\StudyTableWidth}C{0.185\StudyTableWidth}}
\caption{Multivariate Regression Estimates of AI-Human Rating Bias (Spec B)}\label{tab:global_delta_ols_filled_coef_B}\\
\toprule
\textbf{Variable} & \textbf{Claude\newline Coef. (SE)} & \textbf{DeepSeek\newline Coef. (SE)} & \textbf{Gemini\newline Coef. (SE)} & \textbf{GPT\newline Coef. (SE)} \\
\midrule
\endfirsthead
\multicolumn{5}{P{\dimexpr\linewidth-2\tabcolsep-2\arrayrulewidth\relax}}{\textit{Table~\thetable\ (continued)}} \\
\textbf{Variable} & \textbf{Claude\newline Coef. (SE)} & \textbf{DeepSeek\newline Coef. (SE)} & \textbf{Gemini\newline Coef. (SE)} & \textbf{GPT\newline Coef. (SE)} \\
\midrule
\endhead
\endfoot
\endlastfoot
Intercept & 19.14 (1.369)*** & 9.452 (1.262)*** & 0.335 (1.346) & 2.470 (1.358) \\
\multicolumn{5}{P{\dimexpr\linewidth-2\tabcolsep-2\arrayrulewidth\relax}}{\textit{Author characteristics}} \\
Industry involvement & -0.137 (0.039)*** & -0.020 (0.039) & 0.048 (0.041) & 0.126 (0.043)** \\
Top institution & -0.194 (0.043)*** & -0.125 (0.043)** & -0.082 (0.044) & -0.058 (0.048) \\
Female authors & 0.036 (0.059) & -0.041 (0.059) & -0.091 (0.061) & -0.096 (0.067) \\
First author student & 0.018 (0.027) & 0.017 (0.027) & 0.009 (0.028) & -0.002 (0.030) \\
Native English & -0.135 (0.039)*** & -0.029 (0.038) & 0.100 (0.039)* & 0.042 (0.042) \\
N authors & -0.033 (0.005)*** & -0.019 (0.005)*** & -0.011 (0.005)* & -0.019 (0.006)*** \\
\multicolumn{5}{P{\dimexpr\linewidth-2\tabcolsep-2\arrayrulewidth\relax}}{\textit{Paper characteristics}} \\
Formula density & -0.745 (0.100)*** & -0.712 (0.097)*** & -0.372 (0.103)*** & -0.849 (0.109)*** \\
Fig. \& Tab. density & -1.043 (0.245)*** & -0.421 (0.240) & -0.127 (0.254) & -0.083 (0.260) \\
Table info density & 0.113 (0.174) & 0.149 (0.172) & -0.061 (0.174) & -0.174 (0.189) \\
Citation density & -0.081 (0.024)*** & 0.051 (0.024)* & 0.022 (0.026) & 0.111 (0.028)*** \\
Appendix ratio & -2.325 (0.135)*** & -1.749 (0.131)*** & -0.391 (0.145)** & -1.199 (0.148)*** \\
Log body length & -1.511 (0.120)*** & -0.746 (0.112)*** & 0.057 (0.120) & -0.125 (0.121) \\
Title length & 0.013 (0.0004)*** & 0.011 (0.0004)*** & 0.009 (0.0004)*** & 0.009 (0.0004)*** \\
Avg formula length & 1.1e-4 (1.2e-4) & 1.2e-4 (1.2e-4) & -1.4e-4 (1.1e-4) & -3.5e-5 (1.2e-4) \\
Avg sentence length & 0.0009 (0.0008) & 0.0010 (0.0008) & 0.0013 (0.0008) & 0.0015 (0.0008) \\
Abstract length & -8.3e-5 (2.1e-5)*** & -5.3e-5 (2.1e-5)* & -7.9e-5 (2.2e-5)*** & -8.6e-5 (2.4e-5)*** \\
\multicolumn{5}{P{\dimexpr\linewidth-2\tabcolsep-2\arrayrulewidth\relax}}{\textit{Post hoc statistics}} \\
Human rating SD & 0.037 (0.019)* & 0.048 (0.018)** & 0.079 (0.019)*** & 0.046 (0.021)* \\
Human N reviews & -0.028 (0.022) & 0.007 (0.021) & 0.008 (0.022) & 0.005 (0.024) \\
Observations & 7,083 & 7,083 & 7,083 & 7,083 \\
$R^2$ & 0.340 & 0.210 & 0.094 & 0.107 \\
Adj. $R^2$ & 0.338 & 0.208 & 0.091 & 0.104 \\
\multicolumn{5}{P{\dimexpr\linewidth-2\tabcolsep-2\arrayrulewidth\relax}}{\textit{Note:} *** $p<0.001$, ** $p<0.01$, * $p<0.05$. Mean VIF = 1.202, max VIF = 1.956.} \\
\bottomrule
\end{longtable}
\endgroup
\end{landscape}

\clearpage\section{Complete DML Sensitivity Analyses}\label{app:robustness}
\begin{figure}[htbp]
\centering
\includegraphics[width=\linewidth]
{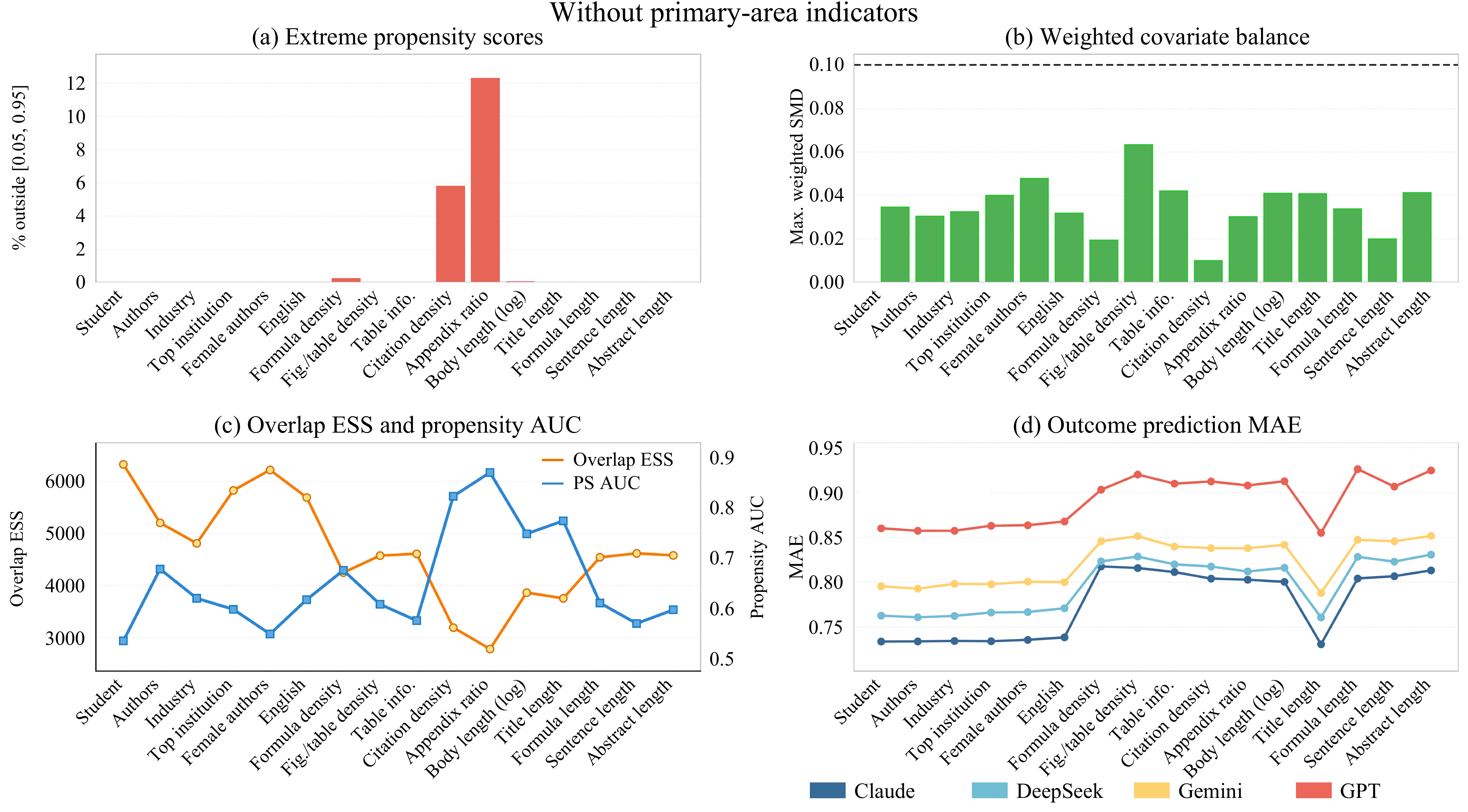}
\caption{Diagnostics without Primary-Area Indicators. Panels report extreme propensities, weighted balance, effective sample size and propensity AUC, and cross-fitted outcome MAE, using the same definitions as the main-specification dashboard.}
\label{fig:rdr_dml_diagnostics_no_area}
\end{figure}

Table~\ref{tab:rdr_dml_ato_no_area} omits primary-area indicators. Table~\ref{tab:causal_512_proxy} uses a 512-token SciBERT proxy based on the title, abstract, and introduction. Both tables report all manuscript and author characteristics. Title-length and formula-density contrasts retain their directions and FDR significance across the four models under both specifications. The remaining contrasts show varying sensitivity to the controls. Changing the proxy changes the information used for adjustment and may change the overlap population. Figure~\ref{fig:rdr_dml_diagnostics_no_area} reports diagnostics for the specification without area indicators.

\begin{landscape}
\begingroup
\small\setstretch{1.05}
\setlength{\tabcolsep}{3pt}
\renewcommand{\arraystretch}{1}
\setlength{\StudyTableWidth}{\dimexpr\linewidth-10\tabcolsep-6\arrayrulewidth\relax}
\begin{longtable}{P{0.26\StudyTableWidth}C{0.185\StudyTableWidth}C{0.185\StudyTableWidth}C{0.185\StudyTableWidth}C{0.185\StudyTableWidth}}
\caption{RDR-DML-ATO Estimates without Primary-Area Fixed Effects}\label{tab:rdr_dml_ato_no_area}\\
\toprule
\textbf{Variable} & \textbf{Claude\newline Contrast (SE)} & \textbf{DeepSeek\newline Contrast (SE)} & \textbf{Gemini\newline Contrast (SE)} & \textbf{GPT\newline Contrast (SE)} \\
\midrule
\endfirsthead
\multicolumn{5}{P{\dimexpr\linewidth-2\tabcolsep-2\arrayrulewidth\relax}}{\textit{Table~\thetable\ (continued)}} \\
\textbf{Variable} & \textbf{Claude\newline Contrast (SE)} & \textbf{DeepSeek\newline Contrast (SE)} & \textbf{Gemini\newline Contrast (SE)} & \textbf{GPT\newline Contrast (SE)} \\
\midrule
\endhead
\endfoot
\endlastfoot
\textit{Author characteristics} &  &  &  \\
First author student & 0.043 (0.024) & 0.014 (0.024) & -0.007 (0.026) & -0.027 (0.028) \\
N authors & -0.081 (0.027)** & -0.058 (0.027) & -0.043 (0.028) & -0.093 (0.030)** \\
Industry involvement & -0.074 (0.027)* & 0.008 (0.028) & 0.043 (0.029) & 0.105 (0.031)** \\
Top institution & -0.092 (0.025)** & -0.081 (0.026)* & -0.072 (0.027)* & -0.081 (0.029)* \\
Female authors & 0.019 (0.024) & -0.017 (0.024) & -0.023 (0.025) & -0.043 (0.028) \\
Native English & -0.066 (0.025)* & -0.027 (0.026) & 0.038 (0.027) & -0.001 (0.029) \\
\textit{Paper characteristics} &  &  &  \\
Formula density & -0.115 (0.032)*** & -0.180 (0.032)*** & -0.143 (0.033)*** & -0.296 (0.035)*** \\
Fig.\ \& Tab.\ density & 0.010 (0.031) & 0.083 (0.031)* & 0.020 (0.032) & 0.120 (0.034)** \\
Table info density & 0.031 (0.030) & 0.040 (0.030) & -0.008 (0.031) & -0.002 (0.034) \\
Citation density & -0.120 (0.037)** & 0.010 (0.038) & 0.017 (0.038) & 0.033 (0.041) \\
Appendix ratio & -0.226 (0.041)*** & -0.166 (0.042)*** & -0.021 (0.042) & -0.153 (0.046)** \\
Log body length & -0.153 (0.035)*** & -0.090 (0.034)* & -0.036 (0.035) & -0.057 (0.038) \\
Title length & 0.909 (0.034)*** & 0.830 (0.034)*** & 0.742 (0.035)*** & 0.755 (0.038)*** \\
Avg formula length & -0.057 (0.031) & -0.068 (0.031)* & -0.012 (0.032) & -0.089 (0.035)* \\
Avg sentence length & 0.062 (0.030) & 0.094 (0.031)** & 0.080 (0.031)* & 0.133 (0.034)*** \\
Abstract length & -0.021 (0.031) & -0.002 (0.031) & -0.040 (0.032) & -0.069 (0.034) \\
\multicolumn{5}{P{\dimexpr\linewidth-2\tabcolsep-2\arrayrulewidth\relax}}{\textit{Notes.} Entries report RDR-DML-ATO estimates from the robustness
specification excluding primary-area fixed effects. Paper-clustered
cross-fitted influence-function standard errors are reported in parentheses.
For nonbinary treatments, the estimand contrasts the upper and lower terciles
of the treatment distribution, with the middle tercile excluded.
Benjamini-Hochberg correction is performed separately within each AI model
and treatment block.
*** $p_{\mathrm{FDR}}<0.001$,
** $p_{\mathrm{FDR}}<0.01$,
* $p_{\mathrm{FDR}}<0.05$.} \\
\bottomrule
\end{longtable}
\endgroup
\end{landscape}

\begin{landscape}
\begingroup
\small\setstretch{1.05}
\setlength{\tabcolsep}{3pt}
\renewcommand{\arraystretch}{1}
\setlength{\StudyTableWidth}{\dimexpr\linewidth-10\tabcolsep-6\arrayrulewidth\relax}
\begin{longtable}{P{0.26\StudyTableWidth}C{0.185\StudyTableWidth}C{0.185\StudyTableWidth}C{0.185\StudyTableWidth}C{0.185\StudyTableWidth}}
\caption{Adjusted Overlap Contrasts Using the Alternative 512-Token Human-Rating Proxy}\label{tab:causal_512_proxy}\\
\toprule
\textbf{Variable} & \textbf{Claude\newline Contrast (SE)} & \textbf{DeepSeek\newline Contrast (SE)} & \textbf{Gemini\newline Contrast (SE)} & \textbf{GPT\newline Contrast (SE)} \\
\midrule
\endfirsthead
\multicolumn{5}{P{\dimexpr\linewidth-2\tabcolsep-2\arrayrulewidth\relax}}{\textit{Table~\thetable\ (continued)}} \\
\textbf{Variable} & \textbf{Claude\newline Contrast (SE)} & \textbf{DeepSeek\newline Contrast (SE)} & \textbf{Gemini\newline Contrast (SE)} & \textbf{GPT\newline Contrast (SE)} \\
\midrule
\endhead
\endfoot
\endlastfoot
\multicolumn{5}{P{\dimexpr\linewidth-2\tabcolsep-2\arrayrulewidth\relax}}{\textit{Author characteristics}} \\
Industry involvement & -0.095 (0.028)** & -0.005 (0.028) & 0.042 (0.029) & 0.102 (0.031)** \\
Top institution & -0.108 (0.025)*** & -0.085 (0.026)** & -0.063 (0.027) & -0.073 (0.029)* \\
Female authors & 0.019 (0.024) & -0.017 (0.025) & -0.025 (0.026) & -0.042 (0.028) \\
First author student & 0.055 (0.024)* & 0.019 (0.025) & -0.012 (0.026) & -0.028 (0.028) \\
Native English & -0.074 (0.026)** & -0.038 (0.026) & 0.031 (0.027) & -0.008 (0.029) \\
N authors & -0.080 (0.028)** & -0.059 (0.027) & -0.041 (0.029) & -0.093 (0.031)** \\
\multicolumn{5}{P{\dimexpr\linewidth-2\tabcolsep-2\arrayrulewidth\relax}}{\textit{Paper characteristics}} \\
Formula density & -0.155 (0.038)*** & -0.205 (0.036)*** & -0.134 (0.037)*** & -0.307 (0.039)*** \\
Fig. \& Tab. density & -0.036 (0.036) & 0.049 (0.033) & 0.020 (0.033) & 0.103 (0.036)** \\
Table info density & 0.043 (0.034) & 0.028 (0.032) & -0.007 (0.032) & -0.015 (0.034) \\
Citation density & -0.558 (0.036)*** & -0.290 (0.035)*** & -0.119 (0.035)** & -0.130 (0.038)*** \\
Appendix ratio & -0.730 (0.036)*** & -0.486 (0.035)*** & -0.190 (0.035)*** & -0.322 (0.038)*** \\
Log body length & -0.489 (0.035)*** & -0.308 (0.034)*** & -0.143 (0.034)*** & -0.169 (0.037)*** \\
Title length & 1.107 (0.035)*** & 0.937 (0.033)*** & 0.766 (0.033)*** & 0.786 (0.036)*** \\
Avg formula length & -0.117 (0.034)** & -0.093 (0.033)** & -0.008 (0.033) & -0.083 (0.035)* \\
Avg sentence length & 0.109 (0.033)** & 0.113 (0.032)*** & 0.073 (0.032)* & 0.130 (0.034)*** \\
Abstract length & -0.033 (0.033) & -0.011 (0.032) & -0.036 (0.032) & -0.079 (0.035)* \\
\multicolumn{5}{P{\dimexpr\linewidth-2\tabcolsep-2\arrayrulewidth\relax}}{\textit{Note:}
Entries report ATO estimates from the primary specification using the
alternative 512-token SciBERT human-rating proxy. Paper-clustered cross-fitted
influence-function standard errors are in parentheses. The specification
includes review-condition controls and primary-area indicators. Nonbinary
treatments contrast the upper and lower terciles. Positive estimates indicate
greater AI-human rating bias. Each row uses its own treatment contrast and overlap population, so its
magnitude has a characteristic-specific interpretation. Benjamini-Hochberg correction is applied separately
within each AI model and characteristic block.
$^{***}p_{\mathrm{FDR}}<0.001$,
$^{**}p_{\mathrm{FDR}}<0.01$,
$^{*}p_{\mathrm{FDR}}<0.05$.
} \\
\bottomrule
\end{longtable}
\endgroup
\end{landscape}

\clearpage\section{Human-Rating Proxy and Estimation Implementation}\label{app:implementation}
\subsection{Full-Text Human-Rating Proxy}
Table~\ref{tab:text_proxy_training_config} specifies the hierarchical human-rating proxy. SciBERT encodes overlapping text segments. The final-layer representation of each segment's first token receives a learned position embedding and enters a document-level Transformer. Masked mean pooling combines these representations for the regression head. Epoch selection and target scaling use the outer-training sample. Predictions summarize full-text information associated with human ratings.

\begingroup
\small\setstretch{1.05}
\setlength{\tabcolsep}{3pt}
\renewcommand{\arraystretch}{1}
\setlength{\StudyTableWidth}{\dimexpr\linewidth-4\tabcolsep-3\arrayrulewidth\relax}
\begin{longtable}{P{0.31\StudyTableWidth}P{0.69\StudyTableWidth}}
\caption{Training Configuration for the Full-Text Human-Rating Proxy}\label{tab:text_proxy_training_config}\\
\toprule
\textbf{Component} & \textbf{Specification} \\
\midrule
\endfirsthead
\multicolumn{2}{P{\dimexpr\linewidth-2\tabcolsep-2\arrayrulewidth\relax}}{\textit{Table~\thetable\ (continued)}} \\
\textbf{Component} & \textbf{Specification} \\
\midrule
\endhead
\endfoot
\endlastfoot
\multicolumn{2}{P{\dimexpr\linewidth-2\tabcolsep-2\arrayrulewidth\relax}}{\textit{Input and prediction target}} \\
Prediction target & Mean human rating, $H_i$ \\
Input text & Title, abstract, and complete main body. Appendices and references are excluded. \\
Document coverage & Complete available main text, with all segments included. \\
\multicolumn{2}{P{\dimexpr\linewidth-2\tabcolsep-2\arrayrulewidth\relax}}{\textit{Text encoding and document architecture}} \\
Text encoder & SciBERT (\texttt{allenai/scibert\_scivocab\_uncased}) \\
Segment length & 512 tokens, including special tokens \\
Segment overlap & 128 content tokens \\
Maximum document-position capacity & 384 segments. Inputs exceeding this capacity trigger an error. \\
Segment representation & Final-layer SciBERT representation of the first token \\
Document positional information & Learned segment-position embeddings \\
Document-level encoder & Two-layer Transformer encoder \\
Attention heads & 8 \\
Feed-forward dimension & $4\times$ the SciBERT hidden dimension \\
Activation & GELU \\
Dropout & 0.10 \\
Document pooling & Masked mean pooling \\
Regression head & Layer normalization, dropout, hidden projection with GELU, dropout,
and scalar output layer \\
\multicolumn{2}{P{\dimexpr\linewidth-2\tabcolsep-2\arrayrulewidth\relax}}{\textit{Out-of-Fold Construction}} \\
Outer cross-fitting & Five paper-level folds \\
Inner validation & 10\% of each outer-training sample \\
Model-selection criterion & Validation MAE \\
Refitting procedure & After epoch selection, the model is reinitialized and trained on the complete
outer-training sample before prediction on the outer-test fold \\
Target scaling & Fold-specific standardization using training-sample mean and standard
deviation \\
Final human-rating proxy & Out-of-fold prediction $\hat H_i$ \\
\multicolumn{2}{P{\dimexpr\linewidth-2\tabcolsep-2\arrayrulewidth\relax}}{\textit{Optimization}} \\
Optimizer & AdamW \\
Loss function & Mean squared error \\
Learning-rate schedule & Linear decay with warmup \\
Peak learning rate & $1\times10^{-5}$ \\
Weight decay & 0.01 \\
Warmup ratio & 0.06 \\
Maximum epochs & 5 \\
Early stopping & Patience of 2 epochs, with a minimum of 2 training epochs \\
Paper batch size & 1 \\
Gradient accumulation & 8 steps \\
Effective paper batch size & 8 \\
Encoder segment microbatch & 16 segments \\
Gradient clipping & Maximum norm of 1.0 \\
Gradient checkpointing & Enabled \\
Numerical precision & BF16 when supported, otherwise FP16 mixed precision \\
\bottomrule
\end{longtable}
\endgroup

\subsection{Orthogonal Score and Clustered Inference}

For each characteristic, $T_i$ identifies the two comparison groups and $X_i$ contains the adjustment variables. Let $e(X_i)=\Pr(T_i=1\mid X_i)$ and $m_t(X_i)=\mathbb E[Y_i\mid T_i=t,X_i]$, where $Y_i=\Delta_i$ is AI-human rating divergence. Hats denote cross-fitted predictions. With $\lambda(e)=e(1-e)$, $\dot\lambda(e)=1-2e$, and $\hat\tau_i=\hat m_{1i}-\hat m_{0i}$, the estimating score is
\begin{align*}
\hat\psi_i&=\hat\tau_i+
\frac{T_i(Y_i-\hat m_{1i})}{\hat e_i}-
\frac{(1-T_i)(Y_i-\hat m_{0i})}{1-\hat e_i},\\
\hat N_i&=\lambda(\hat e_i)\hat\psi_i+
\dot\lambda(\hat e_i)\hat\tau_i(T_i-\hat e_i),\\
\hat D_i&=\lambda(\hat e_i)+\dot\lambda(\hat e_i)(T_i-\hat e_i),\\
\hat\theta_s&=\frac{\mathbb P_n(\hat N_{is})}{\mathbb P_n(\hat D_{is})}.
\end{align*}
Here $\mathbb P_n$ denotes the sample average, and $s$ indexes a sample partition. The first equation adjusts the outcome contrast using the propensity and outcome predictions. The derivative terms in the numerator and denominator account for the dependence of overlap weights on the estimated propensity \citep{wang2025rate}. Outer and inner folds group all observations from a paper together. Table~\ref{tab:rdr_dml_config} specifies the nuisance learners, preprocessing, and weighting. 

\subsubsection{Inference and Diagnostics}
For split $s$, $\hat J_s=\mathbb P_n(\hat D_{is})$ and
\[
\hat\phi_{is}=\frac{\hat N_{is}-\hat\theta_s\hat D_{is}}{\hat J_s},
\qquad \hat\Phi_{gs}=\sum_{i\in\mathcal I_g}\hat\phi_{is}.
\]
Group $\mathcal I_g$ contains the observations from paper $g$, and $\hat\Phi_{gs}$ sums their influence scores. With $G$ papers and $n$ observations, the paper-clustered standard error is
\[
\widehat{\mathrm{SE}}_s=
\left[\frac{G}{G-1}\frac{1}{n^2}\sum_g\hat\Phi_{gs}^2\right]^{1/2}.
\]
Ten partitions yield median estimate $\tilde\theta=\operatorname{median}_s\hat\theta_s$ and
\[
\widetilde{\mathrm{SE}}=
\left[\operatorname{median}_s\left\{
\widehat{\mathrm{SE}}_s^2+(\hat\theta_s-\tilde\theta)^2
\right\}\right]^{1/2}.
\]
Two-sided tests use the normal reference distribution. The corresponding 95\% interval is $\tilde\theta\pm1.96\widetilde{\mathrm{SE}}$. This standard error incorporates sampling uncertainty and variation across partitions.

Diagnostic weights are $w_i=T_i(1-\hat e_i)+(1-T_i)\hat e_i$. Effective sample size is $(\sum_iw_i)^2/\sum_iw_i^2$, which describes how much information remains after weighting. Diagnostics report the share of propensities outside $[0.05,0.95]$ and weighted standardized mean differences. They also report propensity AUC and log loss, and cross-fitted outcome MAE, root mean squared error (RMSE), and $R^2$. Median predictions across partitions summarize the fitted models. These diagnostics assess observed overlap, covariate balance, and prediction performance.

\begingroup
\small\setstretch{1.05}
\setlength{\tabcolsep}{3pt}
\renewcommand{\arraystretch}{1}
\setlength{\StudyTableWidth}{\dimexpr\linewidth-4\tabcolsep-3\arrayrulewidth\relax}
\begin{longtable}{P{0.31\StudyTableWidth}P{0.69\StudyTableWidth}}
\caption{Implementation Configuration for the RDR-DML ATO Estimator}\label{tab:rdr_dml_config}\\
\toprule
\textbf{Component} & \textbf{Specification} \\
\midrule
\endfirsthead
\multicolumn{2}{P{\dimexpr\linewidth-2\tabcolsep-2\arrayrulewidth\relax}}{\textit{Table~\thetable\ (continued)}} \\
\textbf{Component} & \textbf{Specification} \\
\midrule
\endhead
\endfoot
\endlastfoot
\multicolumn{2}{P{\dimexpr\linewidth-2\tabcolsep-2\arrayrulewidth\relax}}{\textit{Estimation design}} \\
Target estimand & Adjusted contrast in the overlap population (ATO) \\
Estimator & ATO-specific rate-doubly-robust DML-2 \\
Outer cross-fitting & 5 folds grouped by paper \\
Repeated sample splitting & 10 independent paper-level partitions \\
Inner ensemble estimation & 3-fold grouped cross-validation within each outer-training sample \\
Outcome regressions & Estimated separately for $T=0$ and $T=1$ \\
\multicolumn{2}{P{\dimexpr\linewidth-2\tabcolsep-2\arrayrulewidth\relax}}{\textit{Propensity-score nuisance function}} \\
Regularized logistic regression & L2 penalty, $C=1$ \\
Histogram gradient boosting & Learning rate $=0.05$, maximum iterations $=250$,
maximum leaf nodes $=15$, minimum samples per leaf $=20$,
L2 regularization $=1$ \\
Extremely randomized trees & 250 trees, minimum samples per leaf $=10$,
maximum features $=\sqrt{p}$ \\
Ensemble objective & Inner-fold log loss \\
\multicolumn{2}{P{\dimexpr\linewidth-2\tabcolsep-2\arrayrulewidth\relax}}{\textit{Conditional Outcome Nuisance Functions}} \\
Regularized linear regression & Ridge regression with $\alpha=1$ \\
Histogram gradient boosting & Learning rate $=0.05$, maximum iterations $=250$,
maximum leaf nodes $=15$, minimum samples per leaf $=20$,
L2 regularization $=1$ \\
Extremely randomized trees & 250 trees, minimum samples per leaf $=10$,
maximum-feature fraction $=0.8$ \\
Ensemble objective & Inner-fold mean squared error \\
\multicolumn{2}{P{\dimexpr\linewidth-2\tabcolsep-2\arrayrulewidth\relax}}{\textit{Ensemble and preprocessing}} \\
Ensemble weights & Nonnegative convex weights constrained to sum to one \\
Numerical covariates & Median imputation followed by standardization \\
Categorical covariates & Most-frequent-category imputation followed by one-hot encoding \\
Repeated-paper weighting & Inverse within-paper observation frequency during nuisance fitting \\
Propensity numerical bound & $[0.001,0.999]$ for numerical stability \\
\multicolumn{2}{P{\dimexpr\linewidth-2\tabcolsep-2\arrayrulewidth\relax}}{\textit{Covariate specification}} \\
Author-characteristic treatment & Human-rating predictor + paper characteristics + review condition +
primary-area fixed effects \\
Paper-characteristic treatment & Human-rating predictor + author characteristics + review condition +
primary-area fixed effects \\
Primary specification & Includes primary-area fixed effects \\
Alternative specification & Excludes primary-area fixed effects \\
\bottomrule
\end{longtable}
\endgroup

\clearpage\section{Prompt Implementation}\label{app:prompt_exact}
This appendix specifies the prompt inputs, scoring scales, review forms, and condition-specific instructions.
\input{Prompt_Specifications.tex}

%% file: Prompt_Specifications.tex
\subsection{Inputs, Scales, and Output Structures}
All ten conditions use integer scores for the supplied manuscript. Soundness, Presentation, and Contribution range from 1 to 4 (poor, fair, good, excellent). Rating ranges from 1 to 10, and Confidence ranges from 1 to 5. Rating distinguishes strong reject (1-2), reject (3-4), borderline reject (5), borderline accept (6), accept (7-8), and strong accept or award-level work (9-10). Confidence ranges from very low to very high. Soundness assesses technical rigor, Presentation clarity, and Contribution novelty and significance. Rating is a holistic judgment.

The V1-V5 form orders Summary, Strengths, Weaknesses, Soundness, Presentation, Contribution, Ethical Considerations, Rating, and Confidence. Scored headings follow the pattern ``Soundness: X: <short label>,'' with an integer and explanation. Tone and depth templates order Summary, Soundness, Presentation, Contribution, Strengths, Weaknesses, Questions, Ethics flag and details, Rating, and Confidence. The forms share scales and differ in section order. The tone templates require every section and scored dimension, including Questions, even when the narrative is brief.

\subsection{Structured Conditions: V1-V5}
\textit{V1: baseline.} The role is ``You are an ICLR reviewer.'' It requests a complete, structured review of the title, abstract, main text, and supplied appendices. It asks for specific strengths and weaknesses, ethical assessment, and evidence-linked scores. The grounding instruction is: ``Do not invent data, experiments, or citations that are not present in the paper.''

\textit{V2: reviewer guidelines.} V2 retains the form and adds checks on evidence, methods, clarity, reproducibility, and balanced assessment. It asks reviewers to check proofs, experiments, ablations, strong baselines, unrealistic settings, and leakage. Missing support should lower Soundness and Rating. Missing implementation details should reduce Presentation and Soundness. Its distinctive instruction includes: ``Do not give high ratings based on vague impressions such as `interesting idea' without solid empirical or theoretical support.''

\textit{V3: conduct and ethics.} V3 adds objectivity, score discipline, explicit weaknesses, uncertainty, and ethics to the reviewer guidelines. Its conduct rule states: ``Do not speculate about the authors' identity, affiliation, seniority, or intentions.'' Unresolved uncertainty is a negative factor for Soundness and Rating. Unaddressed misuse, fairness, privacy, security, or safety risks count as weaknesses. Adequate discussion and mitigation receive positive recognition.

\textit{V4: Area Chair selectivity.} V4 changes the role to Area Chair and requests a rigorous, conservative review. Ratings of 6 or higher require clear novelty, sound methodology, strong evaluation, and reasonable clarity and reproducibility. Ratings of 7 or higher require comprehensive evidence, non-trivial contributions, and no major unresolved methodological or ethical concern. The tie rule is explicit: ``If you are undecided between two adjacent scores (for example, 6 vs 7), choose the lower score.'' It also prohibits topic-based leniency.

\textit{V5: historical reference and explicit adjustment.} V5 combines conservative assessment, modified review guidelines, and a calibration procedure. The reference states: ``Typical mean reviewer rating in large conferences is around 5.0--5.5.'' Scores of 8 or higher are described as rare and appropriate for exceptional work. The material procedure is:
\begin{enumerate}[leftmargin=2em]
\item ``First, form an internal preliminary overall rating based purely on your qualitative assessment (this is not yet reported).''
\item ``Then, adjust this preliminary rating downward:'' ``In most cases, reduce it by approximately 1 point.'' For noticeable weaknesses, the instruction says to ``reduce it by up to 2 points.''
\item Avoid the adjustment only when novelty, technical strength, empirical validation, clarity, and reproducibility are all strong. Use 7 or higher only across a high bar in all major dimensions, and 8 or higher only for exceptional, award-level or field-changing work.
\end{enumerate}
The practical guidance places promising but imperfect papers typically in the 4-6 range and requires the reported Rating and rationale to reflect calibration. V4-V5 therefore compare instruction bundles. Their differences combine historical reference values, downward adjustment, and conservative assessment.

\subsection{Tone and Depth Conditions}
The five tone/depth templates use the separate form and integer scores. They prohibit systematic score shifting by style. Standard, Constructive, Critical, and Quick restrict judgment to the supplied paper and instructions. Domain Expert also permits generally accepted domain knowledge. It excludes external experimental results, unpublished work, and author-specific assumptions.

\textit{Standard} requests balanced, evidence-based, decision-relevant strengths and weaknesses, concise comments, targeted Questions, and agreement between text and scores. Its constraint states: ``Standard style must NOT systematically shift scores.''

\textit{Constructive} requests prioritized, actionable suggestions and respectful, improvement-oriented criticism while disclosing serious problems. Its core instruction is to ``Frame weaknesses as opportunities for improvement.''

\textit{Critical} requests skeptical examination of assumptions, methods, missing evidence, baselines, ablations, and overclaims while acknowledging strengths. It states: ``Critical style must NOT mechanically downshift scores.''

\textit{Domain Expert} requests technical depth, field-aware novelty assessment, hidden assumptions and edge cases, appropriate baselines, and accessible explanations for nonspecialists. Expertise changes commentary emphasis while retaining the scoring scales.

\textit{Quick} preserves all sections and scores while shortening narrative: Summary uses 2-4 sentences. Strengths and Weaknesses use 2-5 points or short sentences. Scores still require a rationale, especially at extremes. The template states: ``Quick style must NOT systematically change scores.''